# Lunar exosphere dynamics at the north pole following Perseid 2009 meteoroid impacts

Alexey A. BEREZHNOY[1], Maria GRITSEVICH[2,3], Ekaterina A. FEOKTISTOVA[1], Markku NISSINEN[4], Yuri V. PAKHOMOV[5], and Vladislav V. SHEVCHENKO[1]

[1] Sternberg Astronomical Institute, Lomonosov Moscow State University, Universitetskij pr. 13, Moscow 11991, Russia

[2] Faculty of Science, University of Helsinki, Gustaf Hällströmin katu 2a, Helsinki, FI-00014, Finland

[3] Institute of Physics and Technology, Ural Federal University, Mira str. 19, 620002 Ekaterinburg, Russia

[4] Vänrikki Stoolin katu 2 A 17, 00100 Helsinki, Finland

[5] Institute of Astronomy, Russian Academy of Sciences, Pyatnitskaya Street 48, Moscow 119017, Russia

**Abstract** Observations of the lunar exosphere provide valuable insights into dynamic processes affecting the Moon, such as meteoroid bombardment. The Chamberlain model was used to estimate zenith column density and temperature of Na atoms on 13/14 August 2009 after the maximum of the Perseid meteor shower. The column density and temperature of Na atoms delivered to the lunar exosphere by slowly-changing processes during the maximum of the Perseid meteor shower on 12/13 August 2009 are also estimated. Using the Chamberlain model and Monte Carlo simulations, expected ratios of line-of-sight column densities of Na atoms at three observed altitudes on 12/13 August 2009 are obtained. We attribute heightened intensities of Na emission lines on 12/13 August 2009 to the onset of the third short-term peak in Perseid activity predicted by celestial mechanics. The best agreement between observations and theoretical models is achieved with a theoretical temperature of 3000 K for impact-produced Na atoms. This third peak of Perseids is estimated to have begun between 23:29 and 23:41 UT on 12 August 2009, lasting about 83 minutes, with a mass flux attributable to the Perseids ranging between $1.6 \times 10^{-16}$ and $5 \times 10^{-16}$ g cm$^{-2}$ s$^{-1}$. Additionally, depletion of Li content compared to Na content in the lunar exosphere is detected. We developed a model predicting Perseid meteoroid stream activity on the Moon, comparing it with performed spectral observations of the lunar exosphere. By modeling 25,000 years of comet 109P/Swift–Tuttle's orbits, we identified 175 cometary trails likely to have generated meteoroids near the Earth and the Moon during the Perseid 2009 meteor shower. Our results reveal annual maxima inducing filament trail structures, one of which aligned closely with the observed peak of the increased Na content in the lunar exosphere.

## 1. Introduction

Meteoroid impacts leave enduring marks on terrestrial planets and their satellites, shaping the landscapes we see today (Bart and Melosh, 2010; Christou et al., 2014; Schmieder and Kring, 2020; Kreslavsky et al., 2021). These impacts influence the surfaces, atmospheres and exospheres of celestial bodies. When a meteoroid collides with a planetary surface, atmosphere or moon, it sets off a series of physical and chemical changes (Gritsevich et al., 2012; Silber et al., 2018). Beyond mere surface alteration, these impacts have far-reaching effects on atmospheric properties and dynamics (Plane et al., 2007; Pellinen-Wannberg et al., 2014; Silber et al., 2017; Vierinen et al., 2022). On planets like such as Mars, where the atmosphere is thin yet present material ejected into space by impacts leads to long-term atmospheric loss (Schlichting and Mukhopadhyay, 2018). On planetary bodies without substantial atmospheres, meteoroid impacts produce flashes. Monitoring campaigns for impact flashes on the Moon began after the first detections by Bellot Rubio et al. (1998). Although ground-based observations during meteor showers are valuable (Madiedo et al., 2015), they are limited by observational biases. NASA's Lunar Atmosphere and Dust Environment Explorer (LADEE) mission, with its Lunar Dust Experiment (LDEX) detector, was the first attempt at space-based observations monitored dust particles ejected by meteoroid impacts (Szalay and Horányi, 2016; Popel et al., 2018). However, LDEX could only detect dust particles above a certain size and could not directly observe impact flashes. To address these limitations, future missions such as the European Space Agency's Lunar Meteoroid Impact Observer (LUMIO) are being prepared. LUMIO aims to monitor impacts on the far side of the Moon using CCD cameras (Topputo et al., 2023).

The exosphere of the Moon is a very thin atmosphere-like layer consisting of gases and other particles that are gravitationally bound to the Moon. Resonance emission lines of Na and K atoms in the lunar exosphere have been documented over several decades (Potter and Morgan, 1988; Leblanc et al., 2022). Thermal and photon-stimulated desorption are the primary mechanisms for low-energy alkali atom production in the exosphere (Smyth and Marconi, 1995), while meteoroid bombardment and sputtering produce high-energy atoms (Mendillo et al., 1991; Gamborino et al., 2019). Impact vaporization predominantly yields exospheric Na atoms over the Moon's terminator and poles (Sarantos et al., 2010), with increases in Na content during meteor showers (Hunten et al., 1998; Verani et al., 1998, Berezhnoy et al., 2014). Significant increases of Na and K column densities on daily timescales have been detected during meteor shower peaks (Colaprete et al., 2016).

Study of meteoroid impacts on terrestrial planets and their satellites presents an opportunity to examine the influx of meteoroids at varying heliocentric distances through dedicated observations (Jenniskens, 2006; Christou et al., 2019; Christou and Gritsevich, 2024). The monthly and yearly variations of mass flux of meteoroids impacting the Moon was studied by Pokorný et al. (2019). Neslušan and Hajduková (2024) investigated the potential impact of meteoroids from comet 109P/Swift-Tuttle's stream on terrestrial planets beyond Earth, identifying meteor showers at Venus and Mars with similar radiant positions, though none as active as the Perseids at Earth. They predict that although some meteoroids from a filament of 109P's stream should impact Mercury's surface, the resulting shower should be undetectable from Earth. Our preliminary analyses of observations of the lunar exosphere during the Perseid 2009 meteor shower utilized the Chamberlain model (Berezhnoy et al., 2014). This model assumes the lunar exosphere to be symmetric and stationary, which is inconsistent with the observed pronounced latitude-dependency and quick changes of the sodium lunar exosphere (Leblanc et al., 2022).

The Monte Carlo approach is a powerful tool for studying meteoroid impacts (Kastinen and Kero, 2017, Moilanen et al., 2021; Gritsevich et al., 2024a) and their effects on non-stationary exospheres (Killen et al., 2010; Mangano et al., 2007) as well as for studies of the solar wind as a source of exospheric species (Vorburger et al., 2024). In this paper, we present the development of Monte Carlo techniques to explore meteoroid bombardment as a source of the lunar exosphere, with a secondary objective of applying these techniques to analysis of performed spectral observations of the lunar exosphere. Our Monte Carlo simulations specifically focus on impact-produced atoms, while the properties of the stationary lunar exosphere are estimated using the Chamberlain model. Additionally, we propose a model to predict the activity of meteoroid streams both on Earth and the Moon, validated against and compared with available observations of the Perseid meteor shower in 2009.

This paper is organized as follows. In Section 2 we estimate properties of impact-produced Na atoms in the lunar exosphere using Chamberlain model. In Section 3 we present a stochastic model of behavior of atoms in the exosphere of the Moon. In Section 4 we estimate the duration and mass flux of the peak of activity of the Perseid meteor shower by finding parameters producing the best fit between the Chamberlain model and our Monte Carlo simulations of impact-produced exospheric Na atoms. In Section 5 we use celestial mechanics to predict the activity of the Perseid meteor shower on the Moon.

## 2. Observed lunar exosphere properties during the Perseid 2009 meteor shower

Spectral observations of the lunar exosphere in Na resonance doublet D1 (5895.9 Å) and D2 (5890.0 Å) lines were performed on 12/13 and 13/14 August 2009 during maximum of the Perseid meteor shower using the echelle spectrograph MMCS (Multi Mode Cassegrain Spectrometer) installed on the 2-m Zeiss telescope (Terskol branch of the Institute of Astronomy of the Russian Academy of Sciences, Kabardino-Balkaria, Russia). The spectrograph with an angular slit size of 10"x2" was set to the mode with the spectral resolution of $\lambda/\Delta\lambda$=13500. The CCD size of 1245x1152 pixels was used for registration of 31 spectral orders in the range between 3720 and 7526 Å. Typical signal-to-noise ratio of the spectra S/N=50 at the position of Na D1 and D2 lines during exposure time of 1800 s. Six echelle spectra were obtained at three angular distances of 50", 150", and 250" from the lunar limb near the north pole of the Moon as it was bombarded by Perseid's meteoroids (see Fig. 1).

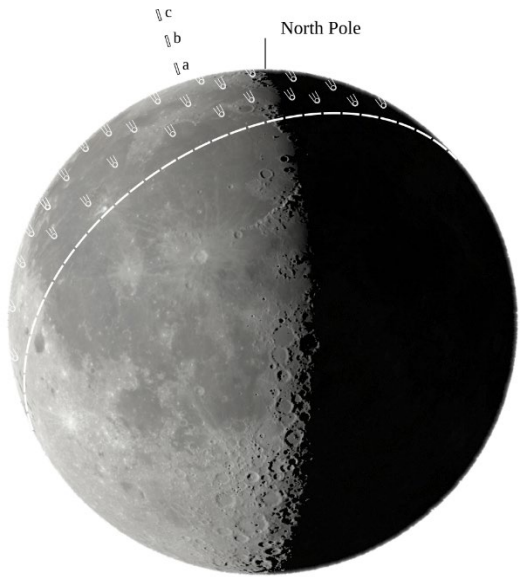

Figure 1. The Moon on 13 August 2009 at 0 UT. Distance between the lunar limb and the slit in positions "a", "b", and "c" is equal to 91, 274, and 456 km, respectively. Position angle of the slit in positions "a", "b", and "c" is 18.9, 18.8 and 18.7 degrees, respectively. The northern part of the lunar surface was bombarded by Perseids. Boundary between regions bombarded and non-bombarded by Perseids is shown by the dotted curve.

To reduce the obtained echelle spectra, we used the Munich Image Data Analysis System (MIDAS). The following steps have been completed: accounting for bias, flat field, cleaning for cosmic rays, detection and extraction the echelle orders, wavelength calibration using spectra of a Fe–Ar hollow-cathode lamp, and flux calibration using the standard star HD214923, which gave us the final spectra in absolute units. These spectra include contributions from the lunar exosphere but are dominated by the solar spectra reflected from the lunar surface and scattered in the terrestrial night sky. To remove the scattered light spectrum, we use the solar spectrum taken with the same instruments when observing the daytime sky. The atmospheric extinction was taken into account in accordance with Tug (1977).

Observations of the lunar exosphere taken during the Perseid 2009 meteor shower (Berezhnoy et al., 2014) were analyzed again using knowledge accumulated since 2014. Specifically, the Chamberlain model was now used to estimate the stationary properties of the lunar exosphere, and the original Monte Carlo approach was then used to study the characteristics of impact-produced metal atoms delivered into the exosphere by Perseid meteoroid impacts.

Knowing observed intensities of Na emission lines, line-of-sight Na column densities (the total number of Na atoms per unit area along a specific path, also known as the line of sight) at three studied altitudes were evaluated (Table 1). Observed line-of-sight column densities $N_{LOS}(X)_{obs}$ of atoms of the element X are calculated as $N_{LOS}(X)_{obs} = 10^6 \, 4\pi I/g$, where $4\pi I$ is the observed intensity of emission of the particular line in Rayleigh and $g$ is the $g$-value (the photon scattering coefficient that is the number of photons one atom scatters per unit time) of the corresponding emission line of atoms of the element X. Surface zenith column densities (these values refers to the vertical integration of the number density of studied atoms specifically along a path from the surface of the Moon to the zenith) and temperatures of Na atoms were calculated by finding minima of least squares between observed line-of-sight Na column densities at three studied altitudes and theoretical line-of-sight Na column densities obtained using the Chamberlain (1963) model of stationary exospheres (see Table 2). Assumptions embedded in the Chamberlain model are: the velocity distribution of species delivered to the exosphere from the surface is Maxwellian, properties of the exosphere such as temperature and zenith column density of species are constant over time, and those properties will be the same over the whole surface of the studied celestial body. The Chamberlain model is therefore not applicable to studies of non-stationary exospheres when properties of the exosphere are quickly changing. However, for stationary exospheres, results from the Chamberlain model and a more realistic Monte Carlo approach agree well at low altitudes (Gamborino et al., 2019).

Table 1. Parameters of spectral observations of Na atoms in the lunar exosphere on 12-14 August 2009. The $g$-values of Na D1 and D2 lines are taken as 0.53 and 0.31 photon atom$^{-1}$ s$^{-1}$, respectively, at temperature of Na atoms of 1700 K (Berezhnoi et al., 2023).

| Time of observations, UT | Distance from the surface $h$, km | Intensity of the Na D2 line $4\pi I(Na)_{obs}$, R | Intensity of the Na D1 line $4\pi I(Na)_{obs}$, R | Intensity of the Na D1+D2 lines $4\pi I(Na)_{obs}$, R | Line-of-sight Na column density $N_{LOS}(Na)_{obs}$, $10^9$ cm$^{-2}$ | | |
|---|---|---|---|---|---|---|---|
| | | | | | based on Na D2 line | based on Na D1 line | based on Na D1+D2 lines |
| Aug. 12, 23:13–23:43 | 91 | 1834±75 | 854±75 | 2688±108 | 3.46±0.14 | 2.76±0.24 | 3.2±0.13 |
| Aug. 12, 23:54–Aug. 13, 0:24 | 274 | 1997±38 | 967±38 | 2964±54 | 3.77±0.07 | 3.12±0.12 | 3.53±0.06 |
| Aug. 13, 0:43–1:13 | 456 | 1720±38 | 804±38 | 2525±54 | 3.25±0.07 | 2.59±0.12 | 3.01±0.06 |
| Aug. 13, 23:22–23:52 | 90 | 1909±38 | 929±38 | 2839±54 | 3.6±0.07 | 3.00±0.12 | 3.38±0.06 |
| Aug. 13, 23:53–Aug. 14, 0:23 | 270 | 1708±38 | 829±38 | 2537±54 | 3.22±0.07 | 2.67±0.12 | 3.02±0.06 |
| Aug. 14, 0:26–0:56 | 450 | 1118±38 | 565±38 | 1683±54 | 2.11±0.07 | 1.82±0.12 | 2.00±0.06 |

Table 2. Observations of Na atoms on 12/13 and 13/14 August 2009.

| Time of observations, UT | Measured Na lines | Surface zenith column density $N_{zen0}$, cm$^{-2}$ | Temperature, K |
|---|---|---|---|
| Aug. 12, 23:13 – Aug. 13, 1:13 | D2 line | (1.56±0.04)×10$^9$ | 11300±3400 |
| Aug. 12, 23:13 – Aug. 13, 1:13 | D1 line | (1.26±0.07)×10$^9$ | 10800±4200 |
| Aug. 12, 23:13 – Aug. 13, 1:13 | D1+D2 lines | (1.45±0.03)×10$^9$ | 10200±1800 |
| Aug. 13, 23:22 – Aug. 14, 0:56 | D2 line | (1.07±0.03)×10$^9$ | 1700±100 |
| Aug. 13, 23:22 – Aug. 14, 0:56 | D1 line | (9.1±0.4)×10$^8$ | 1800±200 |
| Aug. 13, 23:22 – Aug. 14, 0:56 | D1+D2 lines | (1.01±0.02)×10$^9$ | 1700±100 |

During spectral observations of the lunar exosphere, short-term strong activity of the Perseid meteor shower on the Moon was detected on 12/13 August 2009 (Berezhnoy et al., 2014). These observations revealed significant changes only in the intensity of meteoroid bombardment on 12/13 August 2009, while other sources of Na in the lunar exosphere, such as ion sputtering and photon-induced desorption, remained nearly constant (Berezhnoy et al., 2014). The Chamberlain temperature of Na atoms significantly decreased from about 10800 K on 12/13 August 2009 to about 1800 K on 13/14 August 2009 (see Table 2). Typically, observed temperatures of Na atoms change by no more than 30 % at the poles of the Moon on a daily timescale (Killen et al., 2019, 2021). Ion flux on the Moon remained almost constant on 12/13 and 13/14 August 2009 and ion sputtering alone cannot account for the detected rapid and significant changes in Na atom temperature (Berezhnoy et al., 2014). The Chamberlain temperature of Na atoms obtained near the north pole on 12/13 August 2009, about 10800 K, is much higher than the previous Chamberlain temperatures of Na atoms at either pole observed at similar altitudes, lunar phases, and levels of the solar activity (Killen et al., 2021; Kuruppuaratchi et al., 2023). Moreover, partial errors in estimating the Chamberlain temperatures of Na atoms on 12/13 August 2009 are about four times higher than that on 13/14 August 2009 (see Table 2). Based on the discussion above we conclude that the assumption of stationary properties of the Na lunar exosphere during the performed spectral observations on 12/13 August 2009 is invalid.

For this reason, we employed a non-stationary model of the lunar exosphere, taking into account the rapidly changing flux of Perseids on the Moon on 12/13 August 2009 and assuming stationary properties of the exosphere on 13/14 August 2009. The observed line-of-sight column densities of

Na atoms $N_{LOS}(Na)_{obs}$ (see Table 1) on 12/13 August 2009 were approximated as a sum of the line-of-sight column densities of impact-produced atoms $N_{LOS}(Na)_{imp}$ delivered to the exosphere during short-term Perseid's activity and the line-of-sight column densities of Na atoms $N_{LOS}(Na)_{others}$ delivered to the exosphere by other processes such as ion sputtering, thermal and photon-induced desorption, and slowly changing meteoroid bombardment using the formula

$N_{LOS}(Na)_{obs} = N_{LOS}(Na)_{imp} + N_{LOS}(Na)_{others}$ (1).

Surface zenith column density and temperature of Na atoms on 13/14 August 2009 were $10^9$ cm$^{-2}$ and 1800 K, respectively (see Table 2). Column densities and temperatures of Na atoms at the lunar poles can vary by about 30 % on daily timescale (Killen et al., 2019, 2021). For this reason, it was assumed that zenith column densities and temperatures of Na atoms delivered to the exosphere by other processes on 12/13 August 2009 were between $7.5 \times 10^8$ and $1.3 \times 10^9$ cm$^{-2}$ and between 1300 and 2500 K, respectively. Using these assumptions, line-of-sight column densities of Na atoms $N_{LOS}(Na)_{others}$ on 12/13 August 2009 at observed heights were calculated using the Chamberlain model for studied ranges of surface zenith column densities and temperatures. After it, line-of-sight column densities of impact-produced Na atoms $N_{LOS}(Na)_{imp}$ on 12/13 August 2009 were calculated using Eq. (1).

The zenith hourly rate (ZHR) of the Perseid 2009 meteor shower on Earth was reported to be ZHR = 65 on 12 August, 23 UT – 13 August, 1 UT, and ZHR = 25 on 13 August, 23 UT – 14 August 2009, 1 UT (Molau and Kac, 2009). Two activity maxima were detected on Earth on 12 August 2009 at 17 UT (ZHR = 160) and 13 August 2009 at 6 UT (ZHR = 200), which were before and after the observations of the lunar exosphere. Activity peaks on the Moon will be different from those on Earth due to the significant distance between these bodies and the expected presence of small (less than 400,000 km) regions within the streams with higher meteoroid concentrations. As the Perseids were more active on Earth on 12/13 August than on 13/14 August, it was assumed that $N_{LOS}(Na)_{imp} \geq 0$ at all three observed heights on 12/13 August 2009.

Calculated expected Chamberlain line-of-sight column densities of impact-produced Na atoms $N_{LOS}(Na)_{imp}$ are less than $1.3 \times 10^9$ cm$^{-2}$ at 91 km altitude, $(0.9 - 1.9) \times 10^9$ cm$^{-2}$ at 274 km altitude, and $(0.7 - 1.7) \times 10^9$ cm$^{-2}$ at 456 km altitude (see Fig. 2). Calculated expected Chamberlain ratios are between 0.7 and 1.05 for $N_{LOS}(Na, 456\ km)_{imp} / N_{LOS}(Na, 274\ km)_{imp}$ ratio, less than 0.7 for $N_{LOS}(Na, 91\ km)_{imp} / N_{LOS}(Na, 274\ km)_{imp}$ ratio, and less than 0.8 for $N_{LOS}(Na, 91\ km)_{imp} / N_{LOS}(Na, 456\ km)_{imp}$ ratio (see Fig. 3). The next step is to estimate ratios of line-of-sight column densities of impact-produced Na atoms at observed altitudes using a Monte Carlo approach.

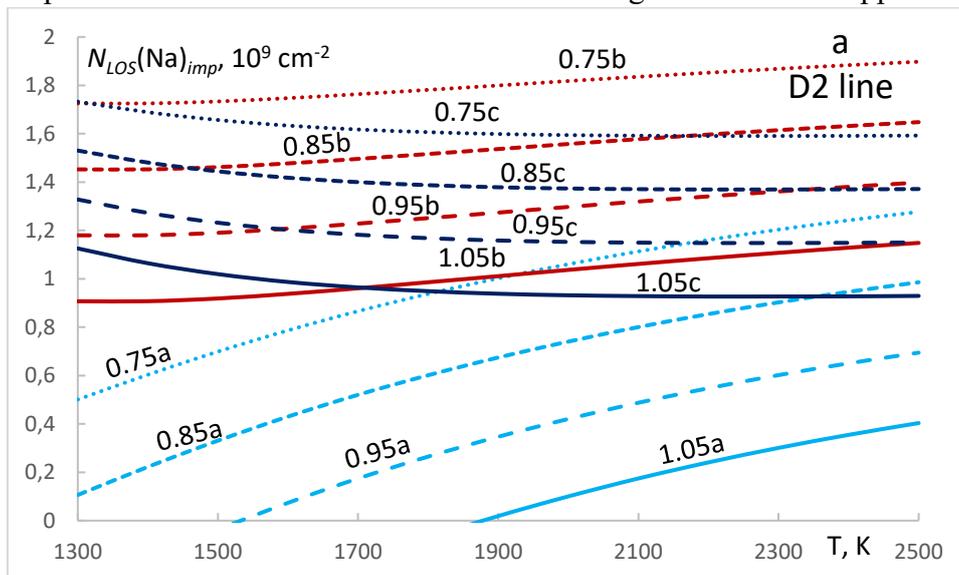

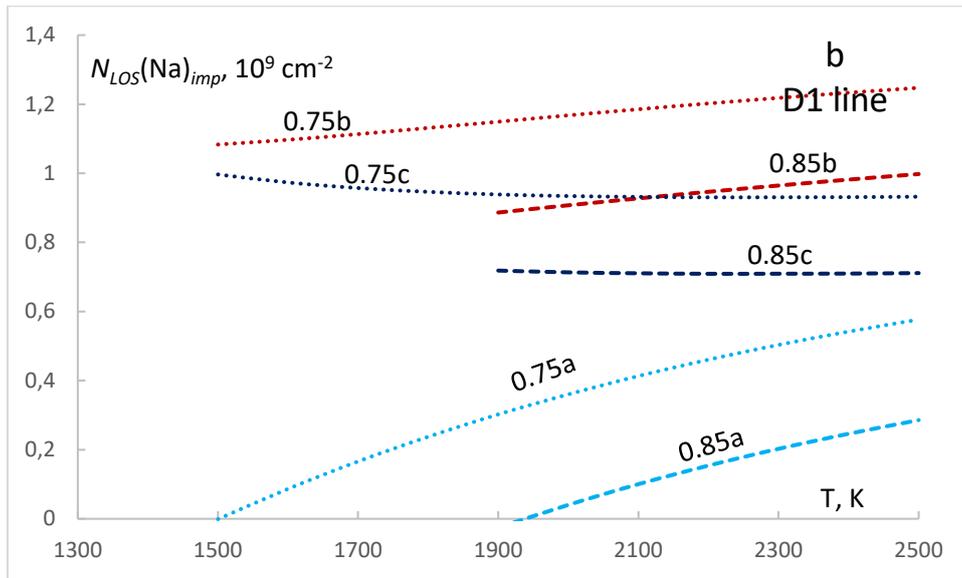

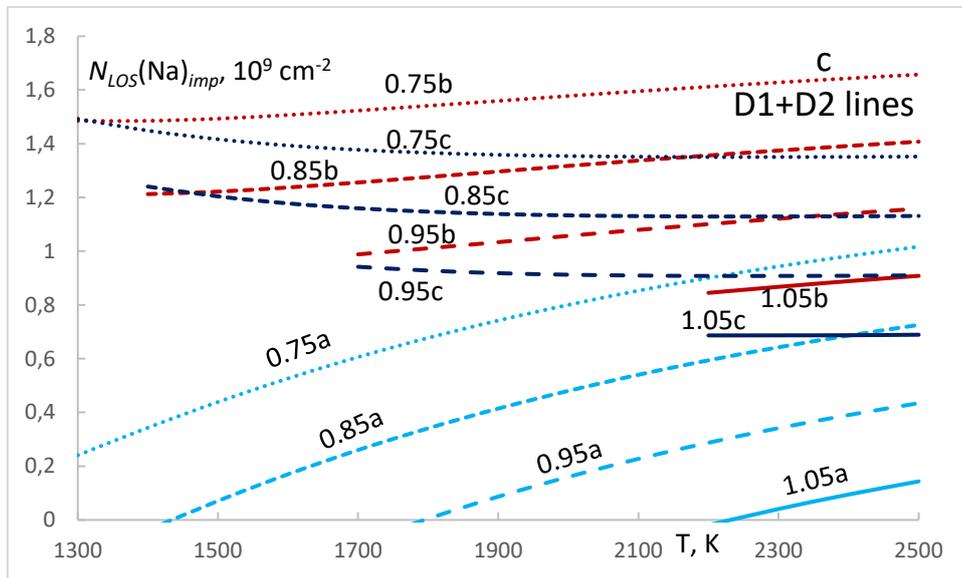

Figure 2. Chamberlain line-of-sight column density of impact-produced Na atoms $N_{LOS}(Na)_{imp}$ at three observed heights on 12/13 August 2009 as a function of temperature and surface zenith column density of Na atoms delivered by other mechanisms $N_{zen0}(Na)_{others}$ based on observations of the Na D2 line (a), D1 line (b), and D1+D2 lines (c). Curves 0.75a, 0.85a, 0.95a, 1.05a are for 91 km height and $N_{zen0}(Na)_{others}$ value equal to $7.5 \times 10^8$, $8.5 \times 10^8$, $9.5 \times 10^8$, and $1.05 \times 10^9$ cm$^{-2}$, respectively. Curves 0.75b, 0.85b, 0.95b, 1.05b are for 274 km height and $N_{zen0}(Na)_{others}$ value equal to $7.5 \times 10^8$, $8.5 \times 10^8$, $9.5 \times 10^8$, and $1.05 \times 10^9$ cm$^{-2}$, respectively. Curves 0.75c, 0.85c, 0.95c, 1.05c are for 456 km height and for $N_{zen0}(Na)_{others}$ value equal to $7.5 \times 10^8$, $8.5 \times 10^8$, $9.5 \times 10^8$, and $1.05 \times 10^9$ cm$^{-2}$, respectively.

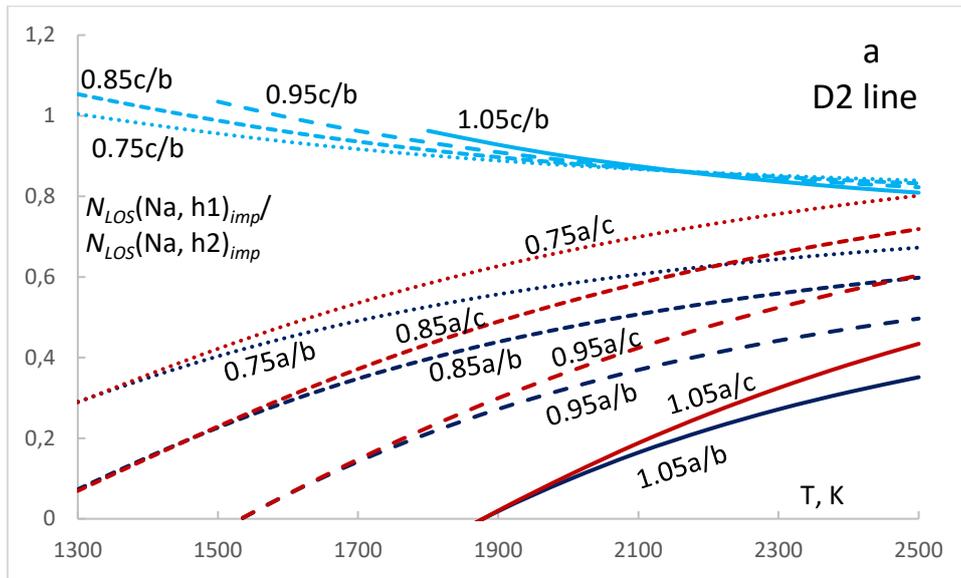
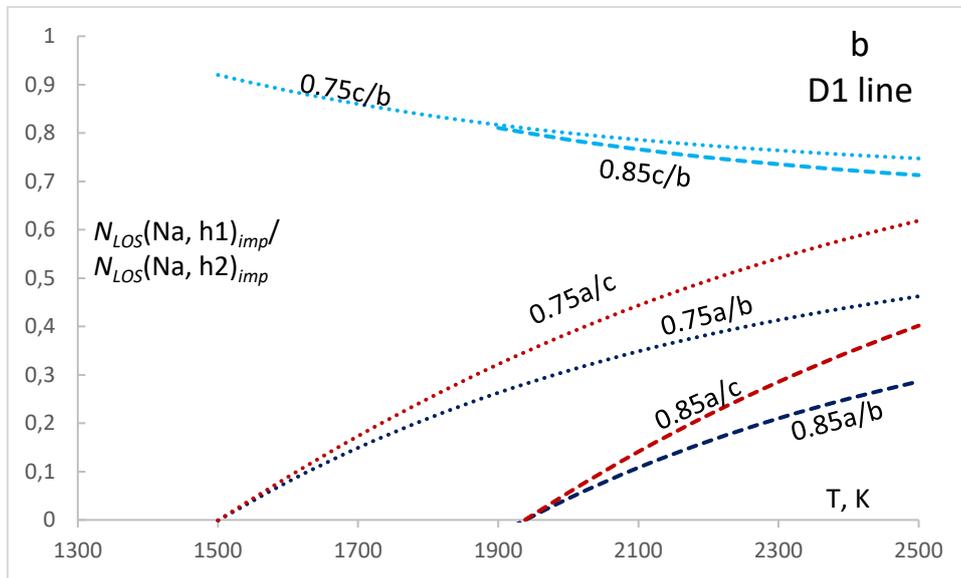
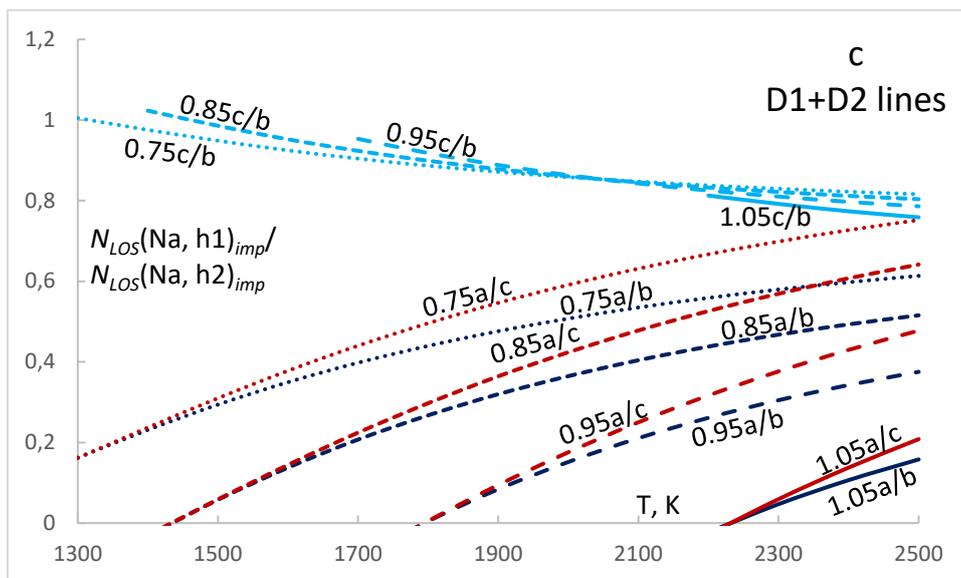

Figure 3. Expected ratios of Chamberlain line-of-sight column densities of impact-produced Na atoms $N_{LOS}(Na)_{imp}$ at two observed heights $h1$ and $h2$ on 12/13 August 2009 as a function of temperature and surface zenith column density of Na atoms delivered by other mechanisms

$N_{zen0}(Na)_{others}$ based on observations of the Na D2 line (a), D1 line (b), and D1+D2 lines (c). Curves 0.75a/b, 0.85a/b, 0.95a/b, 1.05a/b are for $N_{LOS}(Na,\ 91\ km)_{imp}$ / $N_{LOS}(Na,\ 274\ km)_{imp}$ ratio and $N_{zen0}(Na)_{others}$ value equal to $7.5\times10^8$, $8.5\times10^8$, $9.5\times10^8$, and $1.05\times10^9$ cm$^{-2}$, respectively. Curves 0.75a/c, 0.85a/c, 0.95a/c, 1.05a/c are for $N_{LOS}(Na,\ 91\ km)_{imp}$ / $N_{LOS}(Na,\ 456\ km)_{imp}$ ratio and $N_{zen0}(Na)_{others}$ value equal to $7.5\times10^8$, $8.5\times10^8$, $9.5\times10^8$, and $1.05\times10^9$ cm$^{-2}$, respectively. Curves 0.75c/b, 0.85c/b, 0.95c/b, 1.05c/b are for $N_{LOS}(Na,\ 456\ km)_{imp}$ / $N_{LOS}(Na,\ 274\ km)_{imp}$ ratio and for $N_{zen0}(Na)_{others}$ value equal to $7.5\times10^8$, $8.5\times10^8$, $9.5\times10^8$, and $1.05\times10^9$ cm$^{-2}$, respectively.

3. **Monte Carlo simulations of the behavior of species in the lunar exosphere**

Properties of impact-produced atoms in the lunar exosphere were modeled using a Monte Carlo approach. The start location is specified as a point on the Moon's surface with coordinates ($\phi_0$, $\lambda_0$). An isotropic angular distribution of impact plume materials released to the exosphere was used (Killen et al., 2010), with the sub radiant point for the Perseid meteor shower (the place on the lunar surface where the radiant of the shower is at the zenith) chosen as the launch site with the highest probability. The probability of collisions at any point is proportionate to the cosine of the angular separation ($\alpha$) of that point from the sub radiant point. Perhaps obviously, the probability of collisions is 0 if $\alpha$ is between 90 and 180 degrees. The initial temperature of impact-produced Na atoms was taken to be equal to 3000 K (Sarantos et al., 2008; Berezhnoy, 2013). The launch direction $\eta$ and take-off angle $\theta$ are determined randomly. As it was obtained during preliminary simulations, relative fraction of atoms of Na, Li, Ca, and Al performing more than 10 jumps on the surface is less than 0.1 %. For this reason, in our further Monte Carlo simulations the number of particle jumps did not exceed 10 and particles were removed from the system after 10$^{th}$ jump.

The velocity at which particles leave the surface of the Moon is determined by the Maxwell distribution function. After leaving, the particles move in orbits with shapes determined by the initial velocity of the particles. Atoms with velocities below the first escape velocity from the Moon move along ballistic trajectories and return to the surface. Atoms with velocities exceeding the first lunar escape velocity move in elliptical orbits ($e<1$, where $e$ is the eccentricity). If the speeds of atoms equal or exceed the second escape velocity for the Moon, they go into parabolic ($e=1$) or hyperbolic orbits ($e>1$). Atoms on elliptical orbits remain in the exosphere for some time. Atoms on parabolic and hyperbolic orbits quickly leave the exosphere of the Moon. Atoms that return to the surface can make up to 10 jumps. The effect of particle acceleration under the influence of the solar radiation pressure was not considered.

The velocity of an atom on a trajectory is defined as (Mirer, 2013):

$$v = \sqrt{\frac{\mu}{p}(1 + 2e\cos u + e^2)} \qquad (2)$$

where $\mu$ is the gravitational parameter, $p$ is the focal parameter, $e$ is the orbital eccentricity, $u$ is the true anomaly. The position of the particle in the orbit is determined by the value of the true anomaly. The change in the true anomaly in each case is chosen so that the length of the orbital section corresponding to this change does not exceed the value of half the visible radius of the spectrometer slit - 2300 m. The time of movement along the trajectory for each atom is made up of the time of passage of all sections of the orbit: $t = \sum_{j}^{N} t_j$.

During flight, a particle can be destroyed as a result of photoionization with the probability $P_f = e^{t/t_f}$, where $t$ is the flight time: $t = \sum_{j}^{N} t_j$, and $t_f$ is the photoionization time at 1 AU ($7.69\times10^2$ s for Al atoms, $3.08 \times 10^3$ s for Ca atoms, $5.15 \times 10^3$ s for Li atoms, and $1.26 \times 10^5$ s for Na atoms (Huebner and Mukherjee, 2015)). At this stage, a random number $\xi$ is generated: if $\xi > P_f$, the particle will be destroyed. If the particle's velocity is less than the first cosmic velocity for the Moon, it will land on the surface. The landing site coordinates ($\phi_1$, $\lambda_1$) are calculated as in (Mirer, 2013). Next, the surface temperature $T$ in a given location and the sticking coefficient of atoms by the surface (i.e. the ratio of the number of atoms adsorbed by a surface to the total number of atoms that impinge upon that surface) are calculated. For calculations of the surface temperature, it was assumed that the surface is flat with albedo equal to 0.11 (Vasavada et al., 1999) while the height of the Sun above the horizon was calculated as in (Hayne and Aharonson, 2015). The sticking coefficient decreases with increasing temperature:

$$b = a_k e^{cT} \quad (3)$$

Based on experimentally obtained values of the sticking coefficient for collisions of Na atoms with $SiO_2$ surface at different temperatures (Yakshinskiy and Madey, 2005), the coefficients of Eq. (3) are estimated as $a_k = 1.51 \pm 0.12$ and $c = -0.004 \pm 0.0005$. The energy of an atom rebounding from the surface is determined by the following relation (Shemansky and Broadfood, 1977)

$$\alpha = (E_0 - E_2)/(E_0 - E_1) \quad (4)$$

where $a=0.62$ is the accommodation coefficient for Na atoms (Hunten et al., 1988), $E_0$ is the energy of atoms colliding with the surface, $E_1$ is the energy of surface atoms, and $E_2$ is the energy of an atom leaving the surface.

We represented the line of sight as a straight line, passing through two points (observer and observed point in the exosphere) and expressed using lunar-centric coordinates. For each particle jump and for each point on the trajectory the distance between that particle and the line of sight was calculated. The 2"×10" slit used during observations of the lunar exosphere in August 2009 (Berezhnoy et al., 2014) was modelled by a slit with radius equal to 2.52" corresponding to $r_{slit} = 4690$ m at the distance between Earth and the Moon. The particle is assumed to cross the line of sight if the distance between the particle and the line of sight is less than $r_{slit}$. One particle can cross the line of sight several times. During these simulations relative fraction of atoms $f_{slit}(X, t, h)$ crossing the slit per minute as a function of time $t$ since impact event (see Fig. 4a) and average duration of stay of emitted atoms $t_{slit}(X, h)$ on the slit at the height $h$ were determined. Relative fraction of atoms X crossing the slit during obtaining of each spectrum $f_{slit\_obs}(X, h)$ is calculated as $f_{slit\_obs}(X, h) = \sum_{t_1}^{t_2} f_{slit}(X, t, h) \Delta t$, where $t_1$ and $t_2$ are the start and end time of obtaining of each spectrum, and $\Delta t = 1$ minute is the time step (see Fig. 4b).

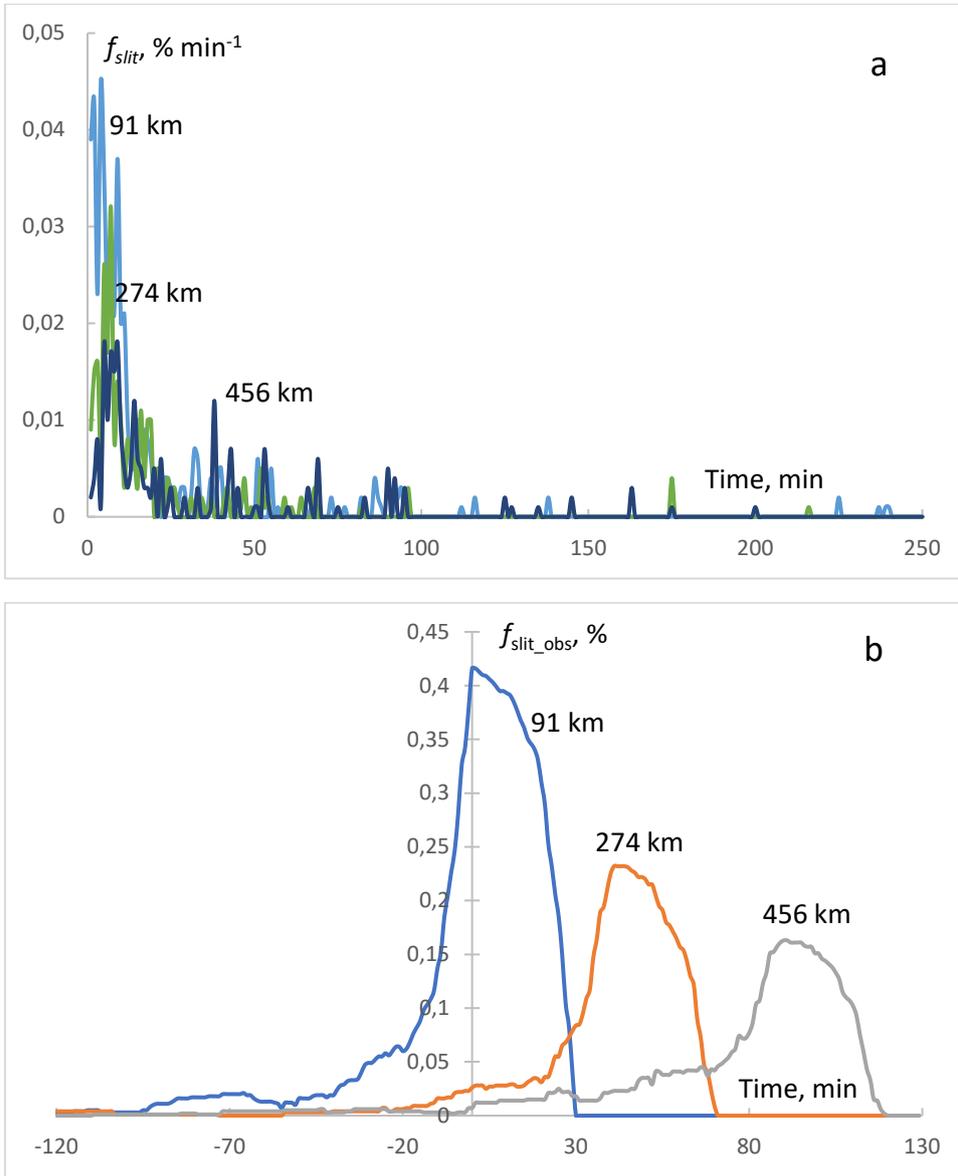

Figure 4. Relative fraction of Na atoms $f_{slit}$ crossing the slit per minute at three observed altitudes as a function of time after release of these Na atoms to the exosphere upon single impact of a meteoroid (a). Relative fraction of Na atoms $f_{slit\_obs}(t)$ crossing the slit during obtaining of spectra of the lunar exosphere on 12/13 August 2009 as a function of time of impact of a single meteoroid, time is equal to 0 on 12 August 2009 at 23:13 UT (b). Initial temperature of Na atoms during its release to the exosphere after impact of a meteoroid is 3000 K. Collisions of Na atoms with the surface leading to decrease of energy of Na atoms in the exosphere and capture of Na atoms by the surface are described by Eqs. 3 and 4. Maximal number of jumps of Na atoms on the surface is 10. Simulations are performed for 100000 particles.

Obtaining $f_{slit\_obs}(X, h)$ and $t_{slit}(X, h)$ values from Monte Carlo simulations, it is possible to estimate theoretical line-of-sight column densities $N_{LOS}(X, h)_{imp}$ of impact-produced atoms X at altitude $h$ in the exosphere as

$N_{LOS}(X, h)_{imp} = M_{imp}\ (F_{target-to-impactor} +1)\ f(X)\ f_{atom}(X)\ f_{unc}(X)\ f_{slit\_obs}(X, h)\ t_{slit}(X, h)\ N_a\ /\ t_{obs}\ S_{slit}\ A_r(X)$ (5),

Where $M_{imp}$ is mass of meteoroids collided with the Moon, $F_{target-to-impactor}$ is the target-to-impactor mass ratio in the gas phase of the impact-produced cloud released to the exosphere, $f(X)$ is the mass fraction of considered element X in the impact-produced cloud, $f_{atom}(X)$ is the ratio of the abundance of atoms of the element X produced directly during impact and by photolysis to the total abundance of X-containing species in the gas phase of the cloud, $f_{unc}(X)$ is the fraction of uncondensed species of element X in the gas phase, $t_{obs}$ = 1800 s is the exposure time of each spectrum, $S_{slit} = \pi r_{slit}^2 = 6.91\times10^{11}$ cm$^2$ is the area of the observed region in the lunar exosphere, $N_a$ is the Avogadro number, and $A_r(X)$ is the atomic mass of atoms of the considered element X.

## 4. Results of the Monte Carlo simulations
### 4.1. Studies of Na atoms

Properties of impact-produced Na atoms in the lunar exosphere were studied as a function of the start time and duration of increased activity of the Perseid meteor shower. It is assumed that mass flux of the Perseid shower during its peak is constant. The peak of the Perseid meteor shower activity was therefore modelled by maintaining a constant time interval between launches of previous and next atoms into the exosphere. The duration of such a peak is determined by the number of atoms considered in each Monte Carlo simulation and the assumed time interval between launches of atoms.

Comparing expected Chamberlain (see Fig. 3) and theoretical Monte Carlo (see Fig. 5) ratios of line-of-sight column densities of Na atoms at different altitudes leads us to conclude that a very short flash of the meteor shower modelled by duration of the shower activity equal to 0 cannot explain the observations because in this case the theoretical ratio of line-of-sight column densities at 456 and 274 km altitudes is too low for any time of start of the peak of the activity of the meteor shower (see Fig. 5). The best agreement between expected Chamberlain and theoretical Monte Carlo ratios at 3000 K is reached for the duration of the peak of the Perseid shower which lasted about 83 minutes and started between 23:29 and 23:41 UT on 12 August 2009 (see Fig. 5).

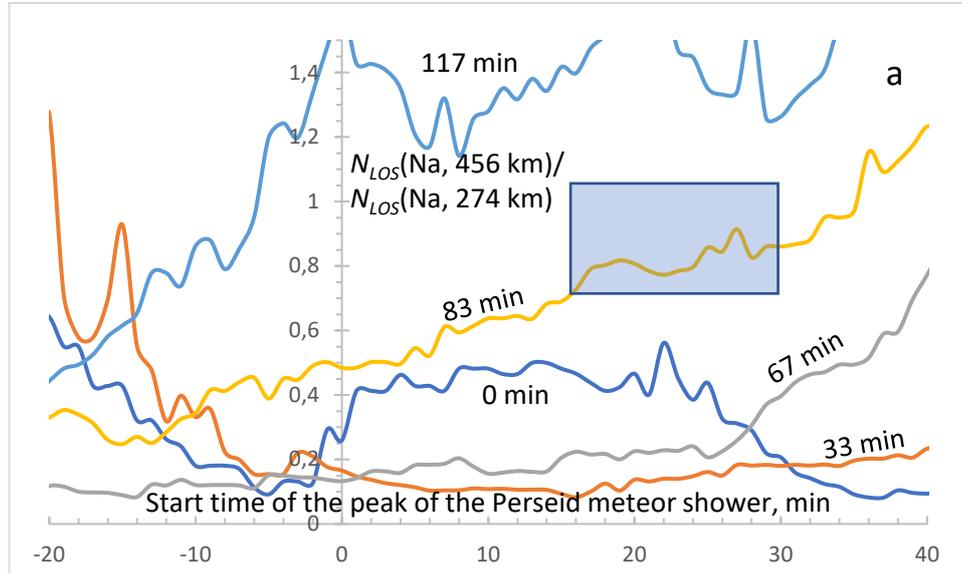

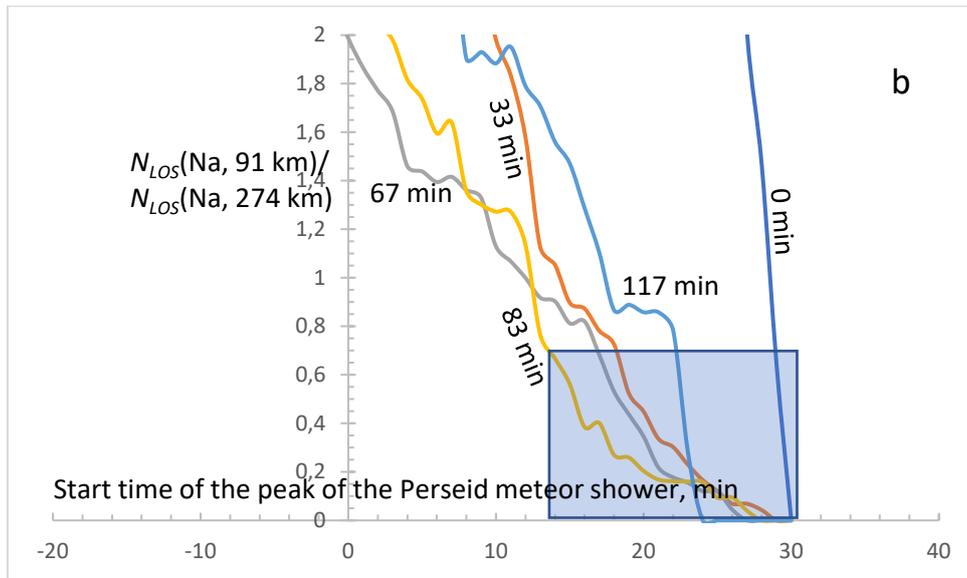

Figure 5. Theoretical Monte Carlo $N_{LOS}$(Na, 456 km)$_{imp}$/$N_{LOS}$(Na, 274 km)$_{imp}$ ratio (a) and $N_{LOS}$(Na, 91 km)$_{imp}$/$N_{LOS}$(Na, 274 km)$_{imp}$ ratio (b) for Na atoms as a function of start time of the peak of the Perseid meteor shower for several durations of this peak. Time is equal to 0 on 12 August 2009 at 23:13 UT. Initial temperature of Na atoms is 3000 K. Number of jumps of Na atoms on the surface is 10, collisions of Na atoms with the surface are described by Eqs. 3 and 4. Calculations are performed for 100000 particles. Rectangles show the expected ranges of start time of the peak of the Perseid meteor shower and Chamberlain $N_{LOS}$(Na, 456 km)$_{imp}$/$N_{LOS}$(Na, 274 km)$_{imp}$ and $N_{LOS}$(Na, 91 km)$_{imp}$/$N_{LOS}$(Na, 274 km)$_{imp}$ ratios obtained from Fig. 3.

The temperature of Na atoms delivered to the lunar exosphere by meteoroid impacts is estimated to be about 3000 K by Sarantos et al. (2008). Due to a presence of Na-containing molecules NaO and NaOH in impact-produced clouds the energy of photolysis-generated Na atoms is larger than the energy of Na atoms formed directly at the quenching of the chemical composition of such clouds (Berezhnoy, 2013). For example, the additional energy of Na atoms obtained during NaO photolysis is 0.17±0.06 eV (Valiev et al., 2020). The temperature of lunar impact flashes is estimated to be in the range between 1300 and 5800 K (Avdellidou and Vaubaillon, 2019). The temperature of vapor clouds produced during high-velocity impact experiments is between 2500 and 5000 K (Eichhorn, 1978). For the reasons discussed above Monte Carlo simulations were also performed for different initial temperatures of Na atoms. At 1000 K theoretical $N_{LOS}$(Na, 456 km)/$N_{LOS}$(Na, 274 km) values are lower than expected Chamberlain values for any start time of the activity of the Perseid shower (see Fig. 6a). The best agreement (estimated as maximal possible interval of start times of the peak of the shower) between theoretical Monte Carlo and expected Chamberlain $N_{LOS}$(Na, 456 km)/$N_{LOS}$(Na, 274 km) and $N_{LOS}$(Na, 91 km)/$N_{LOS}$(Na, 274 km) ratios is achieved for an initial temperature of Na atoms equal to 3000 K (see Fig. 6).

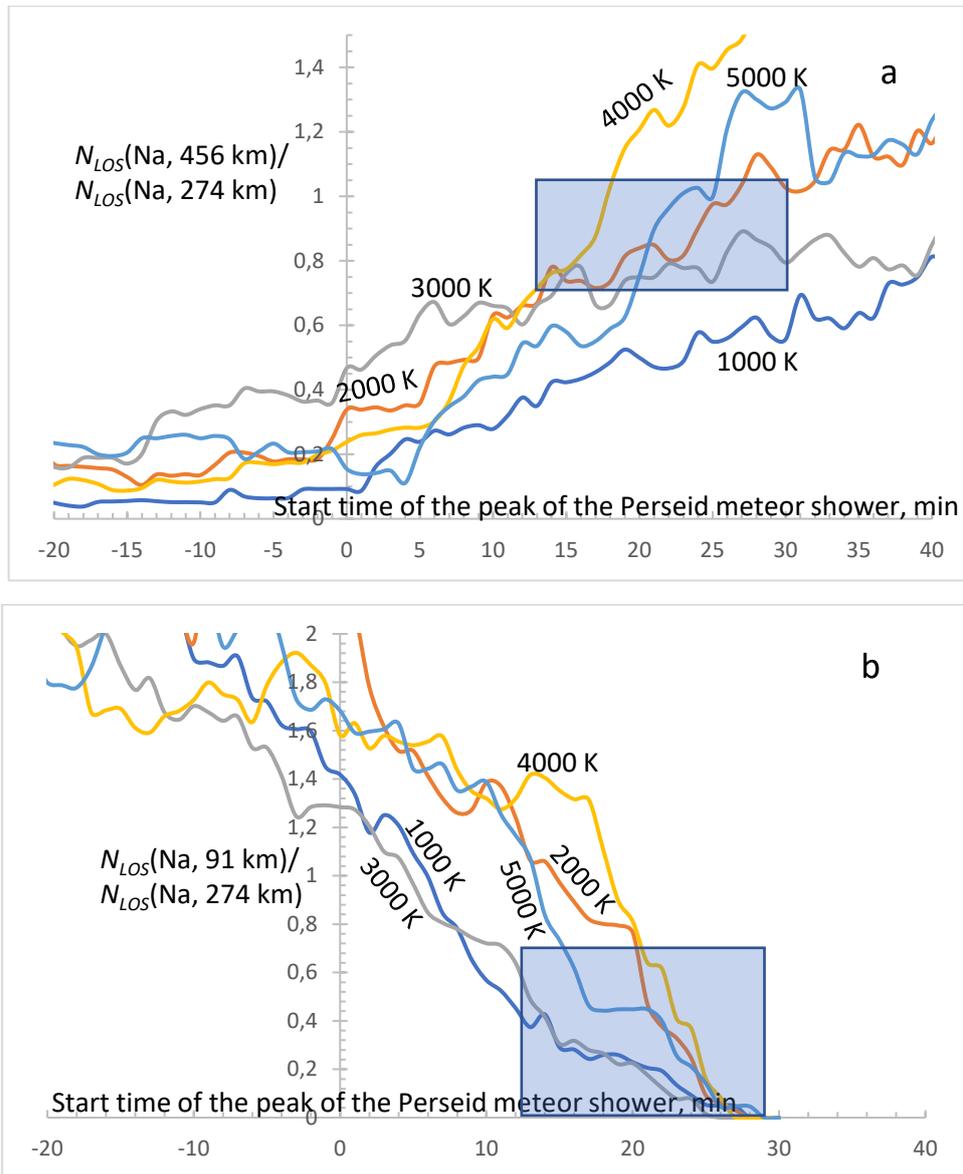

Figure 6. Theoretical Monte Carlo $N_{LOS}$(Na, 456 km)$_{imp}$/$N_{LOS}$(Na, 274 km)$_{imp}$ ratio (a) and $N_{LOS}$(Na, 91 km)$_{imp}$/$N_{LOS}$(Na, 274 km)$_{imp}$ ratio (b) for Na atoms as a function of start time of the peak of the Perseid meteor shower at different initial temperatures of Na atoms. Time is equal to 0 on 12 August 2009 at 23:13 UT. Number of jumps of Na atoms on the surface is 10, collisions of Na atoms with the surface are described by Eqs. 3 and 4. Calculations are performed for 100000 particles. Duration of the peak of the Perseid meteor shower is 83 minutes. Rectangles show the expected ranges of start time of the peak of the Perseid meteor shower and $N_{LOS}$(Na, 456 km)$_{imp}$/$N_{LOS}$(Na, 274 km)$_{imp}$ and Chamberlain $N_{LOS}$(Na, 91 km)$_{imp}$/$N_{LOS}$(Na, 274 km)$_{imp}$ ratios obtained from Fig. 3.

Additional Monte Carlo simulations were performed for different models of interaction of Na atoms with the surface of the Moon (see Fig. 7). The results of these calculations are similar for all models considering loss of energy of Na atoms during collisions with the surface. The $f_{slit\_obs}$ values are similar for the cases of the maximal number of jumps of Na atoms on the surface equal to 1, 3, and 10. It means that change of the maximal number of the possible jumps of Na atoms on the surface from 1 to 10 has only a minor influence on estimation of fraction of Na atoms reaching 456 km altitude.

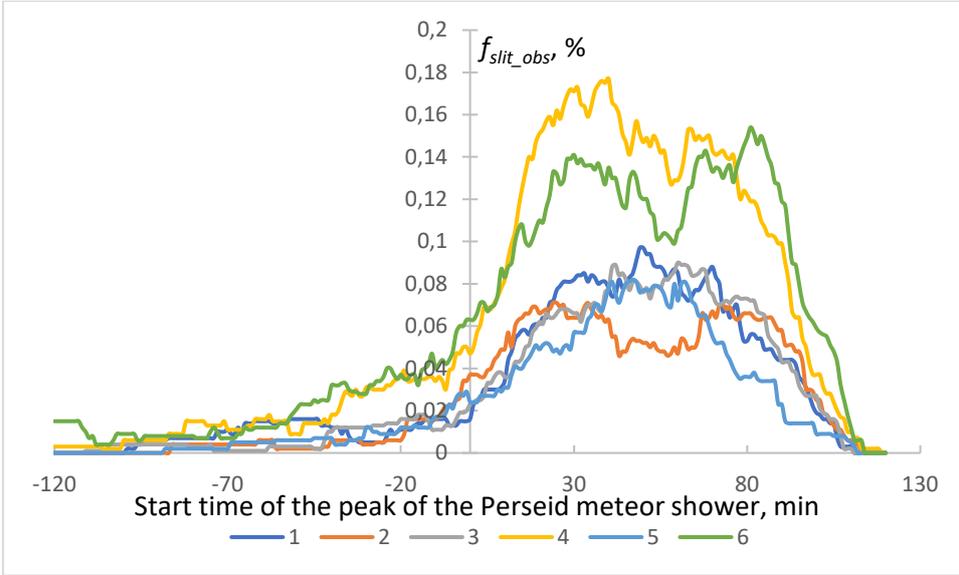

Figure 7. Relative fraction of Na atoms crossing the slit $f_{slit\_obs}$(Na, $t$, 456 km) during obtaining of spectrum of the lunar exosphere at 456 km altitude on 12/13 August 2009 as a function of the start time of the peak of the Perseid meteor shower for different models of interaction of Na atoms with the surface. Time is equal to 0 on 12 August 2009 at 23:13 UT. Initial temperature of Na atoms is 3000 K. Duration of the peak of the Perseid meteor shower is 83 minutes. Simulations are performed for 100000 particles. Curve 1 – number of jumps of Na atoms is 1 (Na atoms are captured after collisions with the surface). Curve 2 – number of jumps of Na atoms is 3, collisions of Na atoms with the surface are described by Eqs. 3 and 4. Curve 3 - number of jumps of Na atoms is 10, collisions of Na atoms with the surface are described by Eqs. 3 and 4. Curve 4 – number of jumps of Na atoms is 10, energy of Na atoms is the same before and after collisions with the surface, Na atoms are not captured by the surface. Curve 5 – number of jumps of Na atoms is 10, energy of Na atoms after collisions is estimated by Eq. 4, Na atoms are not captured by the surface. Curve 6 – number of jumps of Na atoms is 10, sticking coefficient of Na atoms is described by Eq. 3 while energy of Na atoms is the same before and after collisions with the surface.

Calculations with a different number of particles were performed for estimation of statistical errors. It was found that the required accuracy (about 20 %) is achieved for 100000 - 300000 particles.

For start of the peak of activity of the Perseid meteor shower between 23:29 and 23:41 UT on 12 August 2009, temperature of Na atoms equal to 3000 K, and duration of such an activity equal to 83 minutes theoretical number of Na atoms crossing the observed region in the lunar exosphere at 456 km altitude during observations $f_{slit\_obs}$ is 0.056-0.07 % and average duration of stay of Na atoms on the slit at this altitude $t_{slit}$ = 4.2 s. It corresponds to the relative fraction of Na atoms on the slit of about $1.4 \times 10^{-6}$ during obtaining of the third spectrum at 456 km height. Let us assume that in impact-produced clouds the Na content is $f$(Na) = 0.28 wt% and target-to-impactor mass ratio $F_{target\text{-}to\text{-}impactor}$ = 50 (Berezhnoy et al., 2014). It was also assumed that Na is delivered to the lunar exosphere mainly in the form of atoms ($f_{atom}$(Na) = 1) because Na is the main Na-containing compound in impact-produced clouds and condensation of Na-containing species does not occur in impact-produced clouds ($f_{unc}$(Na) = 1) (Berezhnoy et al., 2024), and during its first ballistic flight impact-produced NaO and NaOH molecules are quickly destroyed by solar photons producing atomic Na (Berezhnoy, 2013; Valiev et al., 2020). Using Eq. 5 the aggregate mass of impacting Perseid meteoroids required for agreement between the results of performed Monte Carlo simulations and Chamberlain column densities of impact-produced Na atoms is estimated to be between 100 kg and 230 kg (or mass fluxes between $1.6\times10^{-16}$ and $5\times10^{-16}$ g cm$^{-2}$ s$^{-1}$). These values are significantly higher than that value ($7.7\times10^{-17}$ g cm$^{-2}$ s$^{-1}$) estimated in the earlier paper (Berezhnoy et al., 2014) and the theoretical estimate ($2.8\times10^{-18}$ g cm$^{-2}$ s$^{-1}$) of Hughes and McBride (1989).

*4.2. Studies of behavior of Li, Ca, and Al atoms during meteoroid bombardment of the Moon*

Search for presence of emission lines of atoms of other elements Li, Ca, and Al in the spectra of the lunar exosphere was also performed. Using Eq. 5 of this paper it is possible to estimate theoretical line-of-sight column densities $N_{LOS}(X, h)_{theor}$ of atoms of element X if the parameters of interaction of meteoroids with the Moon are known. The mass of impacting meteoroids was already estimated in Section 4.1 based on the analysis of behavior of Na atoms in the lunar exosphere. Let us assume that for atoms of all considered elements $f_{unc}(X) = f_{atom}(X) = 1$. The $f_{slit\_obs}(X, h)$ and $t_{slit}(X, h)$ values can be estimated during Monte Carlo simulations. The target-to-impactor mass ratio $F_{target-to-impactor}$ was estimated in Section 4.1. Knowing $N_{LOS}(X, h)_{theor}$ and g-values, it is possible to estimate theoretical intensities $4\pi I(X, h)_{theor}$ of emission lines of atoms of other elements.

The depletion factor $Q(X)$ of abundance of atoms of X in the exosphere can be defined as
$Q(X) = 4\pi I(X, h)_{theor}/4\pi I(X, h)_{obs}$ (6),
Where $4\pi I(X, h)_{theor}$ is the theoretical emission intensity of the atoms of X in study at a given altitude estimated by Monte Carlo simulations, and $4\pi I(X, h)_{obs}$ is the upper limit of the emission intensity of the atoms of X in study at a given altitude $h$ at the 3 sigma level. $4\pi I(X, h)_{obs}$ values for emission lines of Li, Al, and Ca atoms are taken from previous observations (Berezhnoy et al., 2014; Berezhnoi et al., 2023). The physical sense of the depletion factor $Q(X)$ is the following. The depletion factor $Q(X) = 1$ if the observed content of atoms of the element X in the exosphere is the same as theoretical one assuming that parameters of interaction of atoms of X and Na with the surface are the same, that element X is released to the exosphere only in the form of atoms ($f_{atom}(X)=1$) and species containing element X do not condense during expansion of the impact-produced cloud ($f_{unc}(X)=1$). In this case the difference in abundances of studied atoms is caused by different mass fraction $f(X)$ of considered elements in the impact-produced cloud as well as by different photo ionization lifetimes of atoms of X and atomic masses $A_r(X)$ affecting also $f_{slit\_obs}(X, h)$ and $t_{slit}(X, h)$ values.

The results of estimation of depletion factors $Q$ are shown in Table 3. For estimation of depletion factors $Q$ for impact-produced atoms (case (g) in Table 3) $f_{slit\_obs}(X, h)$ and $t_{slit}(X, h)$ values for Li, Ca, and Al atoms were calculated for the same initial parameters as for Na atoms (duration of the peak activity is 83 minutes, initial temperature is 3000 K, start time of the peak activity is between 23:29 and 23:41 UT on 12 August 2009, collisions of atoms with the surface are described by Eqs. 3 and 4) giving the best agreement between theory and observations. The case of the constant flux of meteoroids (the case (h) in Table 3) was modelled by achieved the limit of $f_{slit\_obs}(X, h)$ and $t_{slit}(X, h)$ values by increasing duration of activity of the Perseid shower.

Depletion factors $Q$ for Li, Al and Ca atoms are comparable to that obtained using a simple Chamberlain model (Berezhnoy et al., 2014). Depletion of atoms of refractory elements Ca and Al in the lunar exosphere can be explained by condensation of Ca-, Al-containing species in impact-produced clouds ($f_{unc}(X) < 1$) and formation of Ca-, Al-containing compounds stable against photolysis by solar photons ($f_{atom}(X) < 1$) (Berezhnoy, 2013; Berezhnoy et al., 2024). The depletion factor of Li atoms relative to Na atoms in the lunar exosphere is higher than unity according to the observations of Berezhnoi et al. (2023). It means that lithium behavior on the Moon is different from that of sodium. In accordance with thermodynamic calculations it is expected that $f_{unc}(Li) = 1$ because Li is not condensed at the quenching of the chemical composition of impact-produced clouds (Berezhnoy et al., 2024). Main impact-produced LiO, LiOH, and LiCl molecules are quickly form Li atoms during its first ballistic flight in the lunar exosphere (Berezhnoy, 2013; Valiev et al., 2020). For this reason $f_{atom}(Li)$ is also close to unity. Depletion of Li in the lunar exosphere can be explained by higher sticking and accommodation coefficients of Li atoms in comparison with those of Na atoms. These coefficients for Li atoms are still unknown and so additional experimental studies are needed. Different responses of Li and Na on meteoroid impacts are quite possible because the behavior of another alkali element K differs on that of Na during periods of strong meteoroid bombardment of the Moon (Szalay et al., 2016). The Li content in the upper layer of the lunar regolith can be also depleted in comparison with the Na content because the probability of escape from the Moon is higher for lighter Li atoms.

Table 3. Depletion factors $Q$ of Li, Ca, and Al atoms in the lunar exosphere. Initial temperature of atoms of all considered elements is 3000 K. Remarks: a - at 3000 K from Berezhnoi et al. (2023), b – at 0 K from Killen et al. (2022), c – Tera et al., 1970; d – Berezhnoy et al., 2013; e – atoms of Li, Ca, and Al react with the surface in the same way as Na atoms; f – atoms of Ca and Al are captured by the surface after the first ballistic flight; g - impact-produced atoms are delivered to the exosphere during increased activity of the Perseid 2009 meteor shower; h - constant flux of meteoroids and meteoroid bombardment is the unique source of atoms of metals in the exosphere.

| Atom | Li | Ca | Al |
|---|---|---|---|
| Wavelength, Å | 6708 | 4227 | 3962 |
| g-value, photon atom$^{-1}$ s$^{-1}$ | 15.62[a] | 0.5526[b] | 0.0343[b] |
| Content $f(X)$ in impact-produced cloud, wt% | 0.0012[c] | 10.4[d] | 13.1[d] |
| Depletion factor $Q$ relative to Na based on observations on 13 August 2009 at 0:43–1:13 UT[g] | 1.1[e] | 200[e] | 28[e] |
| Depletion factor $Q$ relative to Na based on observations on 12 August 2009 at 23:13–23:43 UT[h] | 0.5[e] | 160[e], 80[f] | 17[e], 11[f] |
| Depletion factor $Q$ relative to Na based on observations of Berezhnoi et al. (2023)[h] | 2[e] | - | - |

## 5. Prediction of activity of the Perseid meteor shower on the Moon

For additional analysis of properties of the Perseid 2009 meteor shower on the Moon and Earth an original celestial mechanics model has been also developed (Gritsevich et al., 2024b). Modeling the dynamics of particles in the stream is analogous to the approach described in Gritsevich et al. (2022). The particle size can be characterized using the $\beta$ parameter, introducing non-gravitational solar radiation pressure disturbances to the particles (Lyytinen, 1999; Lyytinen and van Flandern, 1998; Lyytinen et al., 2001). More specifically, this parameter represents the ratio of radiation pressure to gravitational force acting on a particle (Burns et al., 1979). We conduct simulations for various particle populations, ranging from particulate to fine dust. The $\beta$ parameter is equal to 0.0002, 0.0022, 0.022, and 0.28 for the corresponding population cut-off particle radiuses $r$ equal to 1, 0.1, 0.01, and 0.001 mm, respectively (Lyytinen and Nissinen, 2009; Gritsevich et al., 2022, 2024b).

The non-gravitational forces acting on particles within a comet trail are elucidated in Vaubaillon et al. (2005). Solar radiation pressure results from the Sun's electromagnetic radiation impacting the particles. Additional active forces include the Poynting force and the (diurnal) Yarkovsky-Radzievskii effects, arising from the anisotropy of thermal radiation emitted by the particles.

The simulation of particle escape from the comet involves utilizing the satellite model proposed by Esko Lyytinen (Lyytinen, 1999; Gritsevich et al., 2021). In this model, a comet is characterized by a nucleus accompanied by an orbiting debris cloud system, where the mass of the debris cloud may rival or even exceed that of the primary nucleus. At each perihelion passage, particles are hypothesized to undergo gravitational escape through the L1 and L2 Lagrangian points. These are locations where a particle can remain in a stable position relative to the comet and the Sun, as the gravitational forces and required orbital motion are in balance. This balance of forces allows particles at these points to maintain a stable position relative to the comet, facilitating their possible escape from the comet's gravitational influence. During perihelion passages, when the comet is closest to the Sun, intensified solar radiation and gravitational forces enhance comet's activity. This leads to increased tidal particle release through the L1 and L2 Lagrangian points. The escaped particles sustain and replenish the associated meteoroid streams. An additional mechanism for replenishing meteoroid streams, not examined in this study, involves outbursts or sudden cometary activity. Such events can inject substantial amounts of material into the stream along the comet's orbit (Wesołowski, 2021; Gritsevich et al., 2025).

Smaller particles, primarily meteor-sized, are stable only in the vicinity of cometary nuclei and are propelled away from their stable state mainly by radiation pressure. If ejection speeds are minimal with little dispersion, and orbital dispersion is primarily caused by solar radiation pressure, then radiation pressure acts akin to a decrease in solar gravitation. This results in an increase in the semi-major axis and the period of revolution, proportional to the particle size's radiation pressure/mass ratio. Due to differences in orbital periods between particles, particles form trails after one revolution (Lyytinen and van Flandern, 1998, Gritsevich et al., 2022).

In this work, we investigate the trails left by comet 109P/Swift-Tuttle over a span of 25,000 years. Each trail is named after the year corresponding to a specific perihelion passage of comet 109P/Swift-Tuttle, signifying when that dust trail was formed. Our modeling identifies 175 cometary trails from the comet's past perihelion passages that generate meteoroids in proximity to Earth and the Moon during the Perseids 2009 activity period. This time span reveals an annual maximum that induces filament trail structures (see Fig. 8). The passage of both the Moon and Earth in 2009 through the debris trail of the comet results in the prediction of three distinct activity peaks. These peaks occur due to variations in meteoroid dispersion in the trail and to the influence of specific trails. As Earth and the Moon traverse different parts of the debris trail, they encounter varying concentrations of meteoroids, leading to observable fluctuations in meteor activity. The parameters of these peaks are detailed in Table 4. While the activity of the Perseids 2009 on the Moon is modeled for the first time within this work, the maxima of the Perseid 2009 meteor shower activity on Earth were previously reported (Molau and Kac, 2009; Maslov, 2016) and are in excellent agreement with our modeling results. This agreement reinforces the accuracy and reliability of our modeling approach in predicting meteor shower activity both on Earth and on the Moon.

In Fig. 8, the coordinates are Sun-centered ecliptic coordinates J2000. The positions of Earth and the Moon are obtained from the JPL Horizons System (Giorgini, 2024), and they are Sun-centered ecliptic J2000 coordinates in the International Celestial Reference Frame (ICRF). In the figure, red data points represent Earth, while green data points represent the Moon. The modeled trails, spanning a time frame of 25,000 years, are depicted in yellow. These trail data points are plotted using the time span of 2009.5 – 2009.7 decimal year UT.

The specific characteristics of each peak can be attributed to the distribution of meteoroids within the comet's debris trail. We provide a ZHR value from our model for the relatively coherent 1610 trail peak (1st peak) of the Perseids 2009. The parameters in the model are mean anomaly factor fM = 0.05 showing dispersion of material along the orbit (Mc Naught and Asher, 1999), difference in the semi-major axis of meteoroids' orbits from their parent comet orbit (influenced by ejection velocities and other orbital dynamics (Meng, 2005)), Δa = 0.112 au, and miss distance of the trail from Earth rE-rD = -0.001 au. Our model yields a ZHR value of 89.26 for the Perseids 3rd revolution 1610 trail for the year 2009.

The half-width of the 1610 trail maximum is estimated as described in (Lyytinen et al., 2001). The 1610 trail for the Perseids is comparable to the 4th revolution trail 1866 for the Leonids in 2001. The half-width for that trail was calculated using visual data from the encounters of 1999 and 2000 for the Leonids, with a Δa value of 0.116. The half-width for the Leonids 1866 trail in the year 2001 was calculated to be 43 minutes. Therefore, the half-width for the Perseids 3rd revolution trail 1610 for the year 2009 is estimated to be 50 ± 10 minutes.

Table 4. Parameters of maxima of the Perseid 2009 meteor shower on the Moon and Earth. References: a- Molau and Kac, 2009; b – this work, based on analysis of spectral observations of the lunar exosphere; c – this work, based on the model of spatial structure of the Perseid meteor shower; d – Maslov, 2016.

| Number of the peak | Time of maximum on Earth | Duration of maximum on Earth, minutes | ZHR on Earth | Time of maximum on the Moon | Duration of maximum on the Moon, minutes |
|---|---|---|---|---|---|
| 1 | 8 UT, 12 August 2009[a]; 8:06 UT; 12 August 2009[c, d] | 50±10[c]; 50[a] | 80[a]; 105[c, d] | 4:40 UT, 12 August 2009[c] | - |
| 2 | 17 UT, 12 August 2009[a]; 16:14 UT, 12 August 2009[c, d] | 400[a] | 100[a]; 150[c, d] | 12:48 UT, 13 August 2009[c] | - |
| 3 | 6 UT, 13 August 2009[a]; 6:08 UT, 13 August 2009[c, d] | 150[a] | 140[a]; 173[c, d] | 0:20 UT, 13 August 2009[b]; 2:42 UT, 13 August 2009[c] | 80[b] |

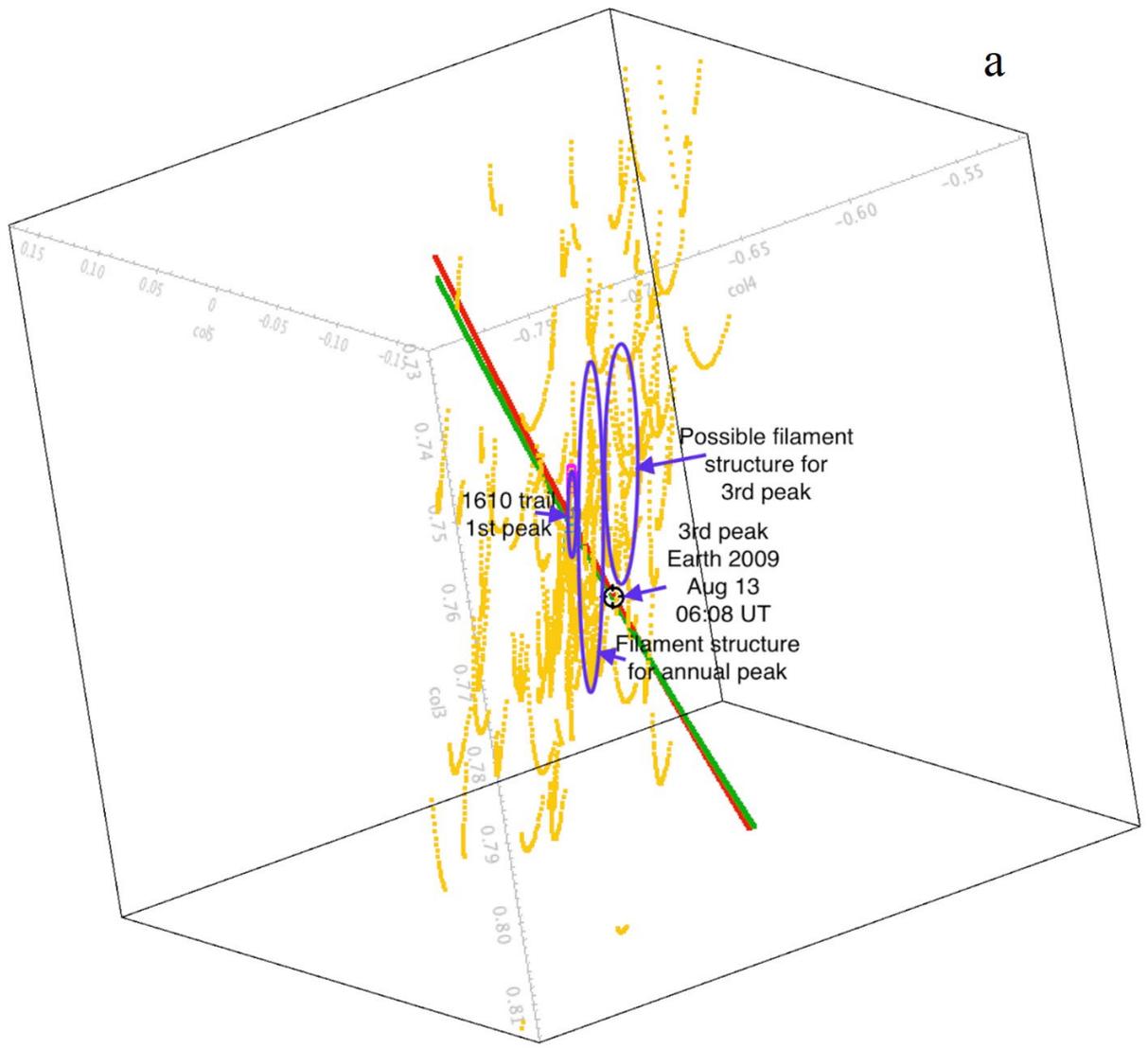

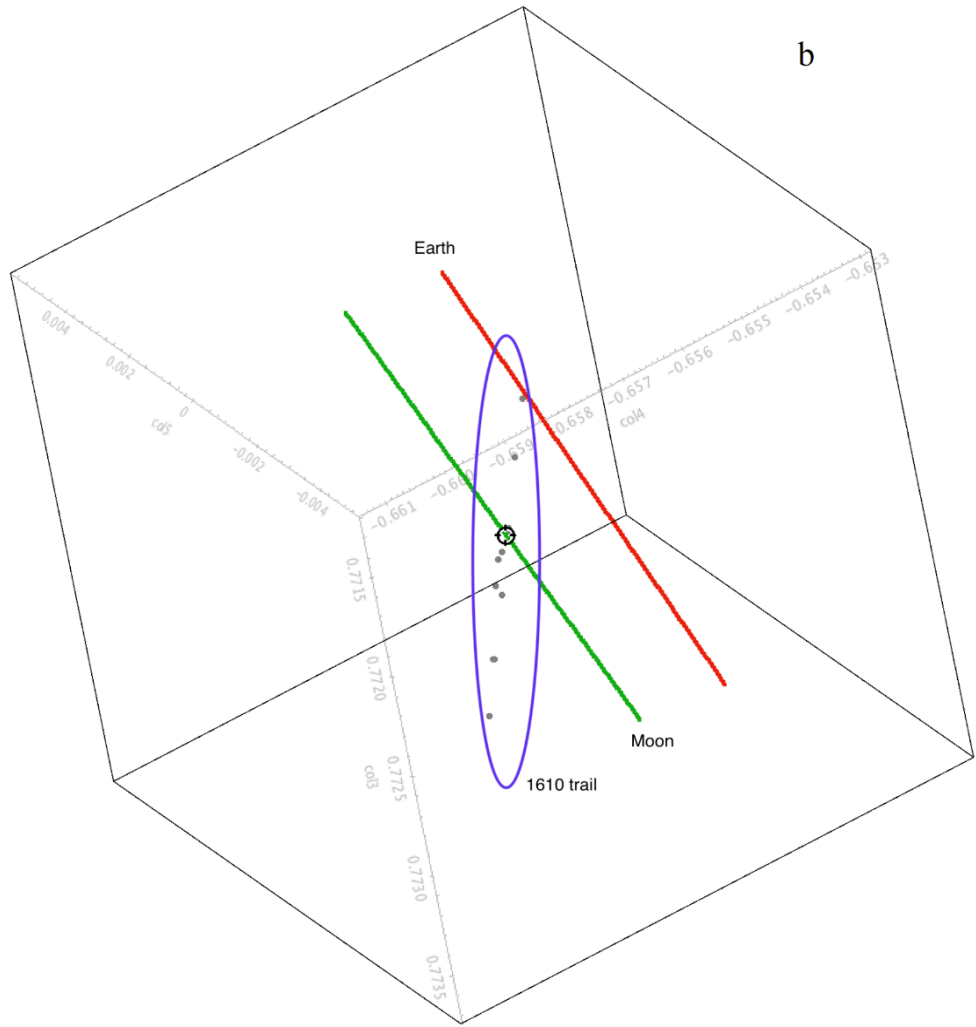

Figure 8. The modeled trails represented in yellow, with the positions of Earth depicted in red and the Moon in green. Coordinates are col3 = X, col4 = Y and col5 = Z. Sun centered ecliptic coordinates J2000 ICRF (a). Zoomed-in picture where particles modeled from the 1610 trail are depicted as circles. The Earth's position is marked in red and the Moon's in green. The coordinates are represented as col3 = X, col4 = Y and col5 = Z, all in Sun-centered ecliptic coordinates of J2000 ICRF (b).

The filament structure observed in Fig. 8 is also sufficient to explain the occurrence of the 3rd peak. Additionally, it is possible that some of the material involved in the 3rd observed peak predates the 25,000-year time span. Factors such as the Poynting-Robertson effect, the gravitational influences of Jupiter and Saturn, have an influence on the dispersion of material in the trail and manifestation of meteor activity on Earth and the Moon. Additional contributing factors capable to cause localized density variations in the debris trail include unaccounted for gravitational disturbances e.g. caused by close encounters with massive objects within or passing through our solar system (Peña-Asensio et al., 2024), resonance effects, more complex cometary activity than accounted for in the model, as well as observational and measurement factors (which might have missed the exact maxima).

Filamentary peaks, characterized by shorter and more localized enhancements in meteor activity within a meteor shower, offer valuable insights into the dynamics of meteoroid streams associated with cometary dust trails. These peaks may arise due to various factors, including debris clumping, resonance effects, and secondary trails not accounted for in cometary activity modeling. Detection of these peaks may also be non-trivial due to observational constraints. Observing, identifying, and modeling these peaks contributes to a more comprehensive understanding of meteoroid stream variability and facilitates more accurate predictions of future meteor shower activity.

In the Perseid 2009 meteor shower, three distinct activity peaks were observed (Maslov, 2016), each with its own explanatory factors related to Earth's orbit proximity to the dust trails left by the comet and variations in meteoroid stream density. We have identified 175 dust trails, situated in close proximity to both Earth and the Moon during the time span of 2009.5 – 2009.7 decimal year UT, which facilitated Perseid 2009 meteor shower activity.

The first peak likely occurred as Earth passed through a denser portion of the debris trail left by the comet, around the solar longitude of 139.624 degrees. Our modeling indicates that this peak was influenced by the debris trail from the comet's perihelion passage in 1610.

The second peak corresponds to the expected annual background maximum, suggesting a more uniform distribution of meteoroids along the debris trail. The slight shift in solar longitude from the first peak may be attributed to Earth traversing different regions of the debris trail.

The third peak is the result of variations in meteoroid stream density, leading to intermittent increases in meteor activity. The identified filament structure of modeled trails contributes to the occurrence of this peak (see Fig. 8). Furthermore, the material comprising this peak might exceed 25,000 years in age (our boundary condition set in modeling). The Poynting-Robertson effect, formally accounted for in the model albeit with a coefficient set to zero, in conjunction with the gravitational influences of Jupiter and Saturn, may shift meteoroids towards Earth and the Moon. While specific trails responsible for the third peak have not been identified within this work, it is plausible that certain individual trails, potentially triggered by heightened activity from a past parent comet outburst, could play a significant role.

The predicted time of the maximum of the third peak on the Moon (2:42 UT on 13 August 2009) occurred after the conclusion of spectral observations of the lunar exosphere at 1:13 UT on 13 August 2009. Consequently, it is not possible to precisely estimate the exact time of the maximum of the third peak of Perseid meteor shower activity on the Moon based solely on the performed spectral observations. There exists a difference of 206 minutes between the theoretical times of maxima of activity of all three peaks on Earth and the Moon.

Assuming that the activity of Perseids on the Moon precedes that on Earth by 206 minutes, it can be inferred that the minimal level of Perseid activity between the second and third peaks occurred around 0 UT on 13 August 2009, on Earth (Molau and Kac, 2009), and approximately at 21:30 UT on 12 August 2009, on the Moon. Therefore, in line with theoretical estimates, the activity of the Perseid shower underwent rapid changes during the conducted spectral observations of the lunar exosphere. Consequently, a steady-state behavior of the Na lunar exosphere was not observed.

## 6. Conclusions

The Chamberlain model was used to estimate the temperature and column density of Na atoms on 13/14 August 2009. Assuming 30 % intra-day variability of these parameters, the temperature and column density of stationary Na lunar exosphere on 12/13 August 2009 is estimated. The Chamberlain model was also used for estimation of expected ratios of line-of-sight column densities of impact-produced Na atoms at three observed altitudes on 12/13 August 2009. After it, Monte Carlo simulations were performed in order to evaluate the Perseid shower as a source of impact-produced Na atoms. The best agreement between theoretical predictions and observations was found for a temperature of impact-produced Na atoms equal to 3000 K. A comprehensive analysis of properties of Na atoms delivered to the exosphere by impacts of Perseids using the original Monte Carlo approach was used for estimate of the mass flux of Perseid meteoroids (between $1.6\times10^{-16}$ and $5\times10^{-16}$ g cm$^{-2}$ s$^{-1}$) responsible for enriching the content of exospheric Na atoms on 12/13 August 2009. The start time (23:29-23:41 UT; 12 August 2009) and duration of the peak activity of the Perseid meteor shower (about 83 minutes) on the Moon were also estimated. The hypothesis of a single impact of a large meteoroid was conclusively rejected because such an impact cannot explain the expected Chamberlain ratio of line-of-sight column densities of impact-produced Na atoms at 456 and 274 km altitudes.

The activity of the Perseid meteor shower on the Moon was also studied using celestial mechanics. This model was validated by the excellent agreement between the observed times of maxima of all three detected main peaks of the Perseids on Earth and the times estimated using the model (see Figure 8 and Table 4). Observations of the lunar exosphere on 12/13 August 2009 were conducted during increasing activity of the Perseid meteor shower before the maximum of the third peak. This supports the theoretical model of Perseid activity on the Moon, in which a short peak of meteoroid bombardment can explain the observations. The model developed for the Perseids on the Moon can be applied to study of other meteor showers on the Moon.

The content of exospheric Li atoms is depleted compared to those of Na atoms, indicating distinct behavior of atoms of these alkali metals on the Moon. Further experimental studies are needed to understand the properties of impact-produced Li atoms in the lunar exosphere, particularly their behavior during collisions with analogues of lunar surface materials and in impact-produced and laser-produced clouds.

Future goals include enhancing the developed Monte Carlo model to incorporate other sources of the lunar exosphere, such as photon-induced desorption, solar wind, thermal desorption, and ion sputtering. Additionally, incorporating impact-produced molecules and considering the photolysis of such molecules in the exosphere is necessary for more comprehensive studies of meteoroid impacts as a source of the lunar exosphere. Investigating the response of the properties of Na and K in the lunar exosphere to rapid changes in parameters of other exospheric sources, such as the solar wind, is also desirable to explain the observed rapid variability of the Moon's exosphere. Such a model could also be utilized for simulating the lunar exosphere during solar eclipses.

## 7. Acknowledgements

The study of A.A.B., E.A.F., and V.V.S. was conducted under the state assignment of Lomonosov Moscow State University. We express gratitude to the Academy of Finland for supporting the project no. 325806 (PlanetS), which facilitated the development of the modelling approaches utilized in this paper. The program of development within Priority-2030 is acknowledged for supporting the research at UrFU. We are grateful to Jim Rowe (The UK Fireball Alliance) for his excellent and proactive help with language corrections, which improved the readability of the paper. We thank Dr. Eloy Peña-Asensio, the Science Operations Center coordinator for ESA's LUMIO mission, for fruitful discussions.

## 8. References

Avdellidou, C., and Vaubaillon, J. 2019. Temperatures of lunar impact flashes: mass and size distribution of small impactors hitting the Moon. *Monthly Notices of the Royal Astronomical Society* **484**(4): 5212-5222.

Bart, G. D., and Melosh, H. J. 2010. Distributions of boulders ejected from lunar craters. *Icarus* **209**(2): 337-357.

Bellot Rubio, L. R., Ortiz, J. L., and Sada, P. V. 1998. Observation and interpretation of meteoroid impact flashes on the Moon. *Earth, Moon, and Planets* **82/83**: 575-598.

Berezhnoi, A., Velikodsky, Yu. I., Pakhomov, Yu. V., and Wöhler, C. 2023. The surface of the Moon as a calibration source for Na and K observations of the lunar exosphere. *Planetary and Space Science* **283**: 105648.

Berezhnoy, A. A. 2013. Chemistry of impact events on the Moon. *Icarus* **226**(1): 205-211.

Berezhnoy, A. A., Churyumov, K. I., Kleshchenok, V. V., Kozlova, E. A., Mangano, V., Pakhomov, Yu. V., Ponomarenko, V. O., Shevchenko, V. V., and Velikodsky, Yu. I. 2014. Properties of the lunar exosphere during the Perseid 2009 meteor shower. *Planetary and Space Science* **96**: 90-98.

Berezhnoy, A. A., Belov, G. V., and Wöhler, C. 2024. Chemical processes during collisions of meteoroids with the Moon. *Planetary and Space Science* **249**: 105942.

Burns, J. A., Lamy, P. L., and Soter, S. 1979. Radiation forces on small particles in the solar system. *Icarus* **40**(1): 1-48.

Chamberlain, J. W. 1963. Planetary coronae and atmospheric evaporation. *Planetary and Space Science* **11**: 901–960.

Christou, A., Oberst, J., Lupovka, V., Dmitriev, V., and Gritsevich, M. 2014. The meteoroid environment and impacts on Phobos. *Planetary and Space Science* **102**: 164-170.

Christou, A. A., Vaubaillon, J., Withers, P., Hueso, R., and Killen, R. 2019. "Extraterrestrial meteors". In *Meteoroids: Sources of Meteors on Earth and Beyond*, edited by G. O. Ryabova, D. J. Asher, and M. G. Campbell-Brown, 119–135. Cambridge: Cambridge University Press. doi:10.48550/437 arXiv.2010.14647.

Christou, A. A., and Gritsevich, M. 2024. Feasibility of meteor surveying from a Venus orbiter. *Icarus* **417**: 116116.

Colaprete, A., Sarantos, M., Wooden, D. H., Stubbs, T. J., Cook, A. M., and Shirley, M. 2016. How surface composition and meteoroid impacts mediate sodium and potassium in the lunar exosphere. *Science* **351**(6270): 249–252.

Eichhorn, G. 1978. Heating and vaporization during hypervelocity particle impact. *Planetary and Space Science* **26**(5): 463-467.

Gamborino, D., Vorburger, A., and Wurz, P. 2019. Mercury's subsolar sodium exosphere: an ab initio calculation to interpret MASCS/UVVS observations from MESSENGER. *Annales Geophysicae* **37**(4): 455-470.Giorgini, J. D. 2024. Horizons On-Line Ephemeris System, Jet Propulsion Laboratory, Pasadena. < https://ssd.jpl.nasa.gov/horizons/ >

Gritsevich, M. I., Stulov, V. P., and Turchak, L. I. 2012. Consequences of collisions of natural cosmic bodies with the Earth's atmosphere and surface. *Cosmic Research* **50**(1): 56–64. http://dx.doi.org/10.1134/S0010952512010017

Gritsevich, M., Nissinen, M., Oksanen, A., Suomela, J., and Silber, E. A. 2022. Evolution of the dust trail of comet 17P/Holmes. *Monthly Notices of the Royal Astronomical Society* **513**: 2201-2214.

Gritsevich, M., Nissinen, M., Moilanen, J., Lintinen, M., Pyykko, J., Jenniskens, P., Brower, J., Moreno-Ibanez, M., Madiedo, J. M., and Rendtel, J. 2021. The miner for out-of-this-world experience. *WGN, Journal of the International Meteor Organization* **49**(3): 52-63.

Gritsevich, M. I., Berezhnoy, A. A., and Nissinen, M. 2024a. Activity of meteoroid streams on the Moon, *55th LPSC conference*, abstract # 2455.

Gritsevich, M., Moilanen, J., Visuri, J., Meier, M. M. M., Maden, C., Oberst, J., Heinlein, D., Flohrer, J., Castro-Tirado, A. J., Delgado-García, J., Koeberl, C., Ferriere, L., Brandstätter, F., Povinec, P. P., Sýkora, I., and Schweidler, F. 2024b. The fireball of November 24, 1970, as the most probable source of the Ischgl meteorite. *Meteoritics and Planetary Science* **59**(7): 1658-1691.

Gritsevich, M., Wesołowski, M., Castro-Tirado, A. J. 2025. Mass of particles released by comet 12P/Pons–Brooks during 2023–2024 outbursts. *Monthly Notices of the Royal Astronomical Society* **538**(1): 470–479.

Hayne, P. O., and Aharonson, O. 2015. Thermal stability of ice on Ceres with rough topography. *Journal of Geophysical Research Planets* **120**(9): 1567-1584.

Huebner, W. F., and Mukherjee, J. 2015. Photoionization and photodissociation rates in solar and blackbody radiation fields. *Planetary and Space Science* **106**: 11-45.

Hughes, D., and McBride, N. 1989. The mass of meteoroid streams. *Monthly Notices of the Royal Astronomical Society* **240**: 73–79.

Hunten, D. M., Morgan, T. H., and Shemansky, D. E. 1988. *The Mercury atmosphere. Mercury.* 562-612. Arizona: University of Arizona Press.

Hunten, D. M., Cremonese, G., Sprague, A. L., Hill, R. E., Verani, S., and Kozlowski, R. W. H. 1998. The Leonid meteor shower and the lunar sodium atmosphere. *Icarus* **136**: 298–303.

Jenniskens, P. 2006. *Meteor showers and their parent comets*. Cambridge: Cambridge University Press.

Kastinen, D., and Kero, J. 2017. A Monte Carlo-type simulation toolbox for Solar System small body dynamics: Application to the October Draconids. *Planetary and Space Science* **143**: 53-66.

Killen, R. M., Potter, A. E., Hurley, D. M., Plymate, C., and Naidu, S. 2010. Observations of the lunar impact plume from the LCROSS event. *Geophysical Research Letters* **37**(23): CiteID L23201.

Killen, R. M., Morgan, T. H., Potter, A. E., Plymate, C., Tucker, R., and Johnson, J. D. 2019. Coronagraphic observations of the lunar sodium exosphere January–June, 2017. *Icarus* **328**: 152-15.

Killen, R. M., Morgan, T. H., Potter, A. E., Bacon, G., Ajang, I., and Poppe, A. R. 2021. Coronagraphic observations of the lunar sodium exosphere 2018–2019. *Icarus* **355**: 114155.

Killen, R. M., Vervack, R. J. Jr., and Burger, M. H. 2022. Updated photon scattering coefficients (g-values) for Mercury's exospheric species. *The Astrophysical Journal Supplement Series* **263**(2): id. 37, 10 pp.

Kreslavsky, M. A., Zharkova, A. Y., Head, J. W., and Gritsevich, M. 2021. Boulders on Mercury. *Icarus* **369**: 114628. https://doi.org/10.1016/j.icarus.2021.114628

Kuruppuaratchi, D. C. P., Oliversen, R. J., Mierkiewicz, E. J., Sarantos, M., and Killen, R. M. 2023. Latitudinal and radial dependence of the lunar sodium exospheric temperature and linewidths. *Icarus* **400**: 115560.

Leblanc, F., Schmidt, C., Mangano, V., Mura, A., Cremonese, G., Raines, J., Jasinski, J. M., Sarantos, M., Milillo, A., Killen, R., Massetti, S., Cassidy, T., Vervack, R., Kameda, S., Capria, M. T., Horanyi, M., Janches, D., Berezhnoy, A., Christou, A., Hirai, T., Lierle, P., and Morgenthaler, J. 2022. Comparative Na and K Mercury and Moon exospheres. *Space Science Reviews* **218**: 2.

Lyytinen, E. 1999. Leonid predictions for the years 1999–2007 with the satellite model of comets. *Meta Research Bulletin* **8**: 33-40.

Lyytinen, E., and Nissinen, M. 2009. Predictions for the 2009 Leonids from a technically dense model. *WGN, Journal of the International Meteor Organization* **37**: 122-124.

Lyytinen, E., Nissinen, M., and van Flandern, T. 2001. Improved 2001 Leonid storm predictions from a refined model. *WGN, Journal of the International Meteor Organization* **29**(4): 110–118.

Lyytinen, E. J., and van Flandern, T. 1998. Predicting the strength of Leonid outbursts. *Earth, Moon and Planets* **82**: 149–166.

Madiedo, J. M., Ortiz, J. L., Organero, F., Ana-Hernández, L., Fonseca, F., Morales, N., and Cabrera-Caño, J. 2015. Analysis of Moon impact flashes detected during the 2012 and 2013 Perseids. *Astronomy and Astrophysics* **577**: A118.

Mangano, V., Milillo, A., Mura, A., Orsini, S., De Angelis, E., Di Lellis, A. M., and Wurz, P. 2007. The contribution of impulsive meteoritic impact vapourization to the Hermean exosphere. *Planetary and Space Science* **55**(11): 1541-1556.

Maslov, M. 2016. Gravitational shifts and the core of Perseid meteoroid stream. *Earth, Moon, and Planets* **117**(2-3): 93-100.

McNaught, R. H., and Asher, D. J. 1999. Leonid dust trails and meteor storms. *WGN, Journal of the International Meteor Organization* **27**: 85-102.

Mendillo, M., Baumgardner, J., and Flynn, B. 1991. Imaging observations of the extended sodium atmosphere of the Moon. *Geophysical Research Letters* **18**: 2097–2100.

Meng, H. 2005. The relationship between the semi-major axis, mass index and age of the Leonid meteor shower. *Monthly Notices of the Royal Astronomical Society* **359**(4): 1433–1436.

Mirer, S.A. 2013. *Space flight mechanics: Orbital motion* // A textbook, Moscow, Rezolit, 270 p. (in Russian)

Moilanen, J., Gritsevich, M., and Lyytinen, E. 2021. Determination of strewn fields for meteorite falls. *Montly Notices of the Royal Astronomical Society* **503**(3): 3337-3350. https://doi.org/10.1093/mnras/stab586

Molau, S., and Kac, J. 2009. Results of the IMO video meteor network — August 2009. *WGN, Journal of the International Meteor Organization* **37**(5): 168-170.

Neslušan, L., and Hajduková, M. 2024. The parts of the meteoroid stream originating in comet 109P/Swift-Tuttle crossing the orbits of Mercury, Venus, and Mars. *Icarus* **415**: 116036.

Pellinen-Wannberg, A. K., Häggström, I., Carrillo Sánchez, J. D., Plane, J. M. C., and Westman, A. 2014. Strong E region ionization caused by the 1767 trail during the 2002 Leonids. *Journal of Geophysical Research* **119**(9): 7880-7888.

Peña-Asensio, E., Visuri, J., Trigo-Rodríguez, J. M., Socas-Navarro, H., Gritsevich, M., Siljama, M., and Rimola, A. 2024. Oort cloud perturbations as a source of hyperbolic Earth impactors. *Icarus* **408**: 115844. https://doi.org/10.1016/j.icarus.2023.115844

Plane, J. M. C., Saiz-Lopez, A., Allan, B. J., Ashworth, S. H., and Jenniskens, P. 2007. Variability of the mesospheric nightglow during the 2002 Leonid storms. *Advances in Space Research* **39**(4): 562-566.

Pokorný, P., Janches, D., Sarantos, M., Szalay, J. R., Horányi, M., Nesvorný, D., and Kuchner, M. J. 2019. Meteoroids at the Moon: Orbital properties, surface vaporization, and impact ejecta production. *Journal of Geophysical Research: Planets* **124**(3): 752-778.

Popel, S. I., Zelenyi, L. M., Golub', A. P., and Dubinskii, A. Yu. 2018. Lunar dust and dusty plasmas: Recent developments, advances, and unsolved problems. *Planetary and Space Science* **156**: 71-84.

Potter, A. E., and Morgan, T. H. 1988. Discovery of sodium and potassium vapor in the atmosphere of the Moon. *Science* **241**: 675–680.

Sarantos, M., Killen, R. M., Sharma, A. S., and Slavin, J. A. 2008. Influence of plasma ions on source rates for the lunar exosphere during passage through the Earth's magnetosphere. *Geophysical Research Letters* **35**(4): CiteID L04105.

Sarantos, M., Killen, R. M., Sharma, A. S., and Slavin, J. A. 2010. Sources of sodium in the lunar exosphere: modeling using ground-based observations of sodium emission and spacecraft data of the plasma. *Icarus* **205**: 364–374.

Schmieder, M., and Kring, D. A. 2020. Earth's impact events through geologic time: A list of recommended ages for terrestrial impact structures and deposits. *Astrobiology* **20**(1): 91–141.

Shemansky, D. E., and Broadfoot, A. L. 1977. Interaction of the surfaces of the Moon and Mercury with their exospheric atmospheres. *Reviews of Geophysics* **15**(4): 491-499.

Silber, E. A., Hocking, W. K., Niculescu, M. L., Gritsevich, M., and Silber, R. E. 2017. On shock waves and the role of hyperthermal chemistry in the early diffusion of overdense meteor trains. *Monthly Notices of the Royal Astronomical Society* **469**(2): 1869-1882. doi: 10.1093/mnras/stx923

Silber, E. A., Boslough, M., Hocking, W. K., Gritsevich, M., and Whitaker, R. W. 2018. Physics of meteor generated shock waves in the Earth's atmosphere – A review. *Advances in Space Research* **62**(3): 489-532. https://doi.org/10.1016/j.asr.2018.05.010

Smyth, W. H., and Marconi, M. L. 1995. Theoretical overview and modeling of the sodium and potassium atmospheres of the Moon. *Astrophysical Journal* **443**: 371–392.

Schlichting, H. E., and Mukhopadhyay, S. 2018. Atmosphere impact losses. *Space Science Reviews* **214**: 34.

Szalay, J. R., and Horányi, M. 2016. Detecting meteoroid streams with an in-situ dust detector above an airless body. *Icarus* **275**: 221-231.

Szalay, J. R., Horányi, M., Colaprete, A., and Sarantos, M. 2016. Meteoritic influence on sodium and potassium abundance in the lunar exosphere measured by LADEE. *Geophysical Research Letters* **43**: 6096–6102.

Topputo, F., Franzese, V., Giordano, C., Massari, M., Pilato, G., Labate, D., Cervone, A., Speretta, S., Menicucci, A., Turan, E., Bertels, E., Vennekens, J., Walker, R., and Koschny, D. 2023. Meteoroids detection with the LUMIO lunar CubeSat. *Icarus* **389**: 115213.

Tug, H. 1977. Vertical extinction on La Silla. *ESO Messenger* **11**: 7–8.

Valiev, R. R., Berezhnoy, A. A., Gritsenko, I. D., Merzlikin, B. S., Cherepanov, V. N., Kurten, T., and Wöhler, C. 2020. Photolysis of diatomic molecules as a source of atoms in planetary exospheres. *Astronomy and Astrophysics* **633**: A39, 12 pp.

Vasavada, A. R., Paige, D. A., and Wood, S. E. 1999. Near-surface temperatures on Mercury and the Moon and the stability of polar ice deposits. *Icarus* **141**: 179–193.

Vaubaillon, J., Colas, F., and Jorda, L. 2005. A new method to predict meteor showers. I. Description of the model. *Astronomy and Astrophysics* **439**(2): 751-760.

Verani, S., Barbieri, C., Benn, C., and Cremonese, G. 1998. Possible detection of meteor stream effects on the lunar sodium atmosphere. *Planetary and Space Science* **46**(8): 1003–1006.

Vierinen, J., Aslaksen, T., Chau, J. L., Gritsevich, M., Gustavsson, B., Kastinen, D., Kero, J., Kozlovsky, A., Kværna, T., Midtskogen, S., Näsholm, S. P., Ulich, T., Vegum, K., and Lester, M. 2022. Multi-instrument observations of the Pajala fireball: Origin, characteristics, and atmospheric implications. *Frontiers in Astronomy and Space Sciences* **9**: 1027750. https://doi.org/10.3389/fspas.2022.1027750

Vorburger, A., Fatemi, S., Carberry Mogan, S. R., Galli, A., Liuzzo, L., Poppe, A. R., Roth, L., and Wurz, P. 2024. 3D Monte-Carlo simulation of Ganymede's atmosphere. *Icarus* **409**: 115847.

Wesołowski, M. 2021. The influence of the size of ice–dust particles on the amplitude of the change in the brightness of a comet caused by an outburst. *Monthly Notices of the Royal Astronomical Society* **505**(3): 3525–3536.

Yakshinskiy, B. V., and Madey, Th. E. 2005. Temperature-dependent DIET of alkalis from $SiO_2$ films: Comparison with a lunar sample. *Surface Science* **593**: 202–209.